\documentclass[a4paper,aps,prd,twocolumn,eqsecnum,amsmath,amssymb,nofootinbib,superscriptaddress]{revtex4-2}
\usepackage{dcolumn}
\usepackage{bm}
\usepackage{graphics}
\usepackage{graphicx}
\usepackage{epsfig}
\usepackage{booktabs,multirow}
\usepackage{hhline}
\usepackage{slashed}
\usepackage{rccol}
\usepackage{mathrsfs}
\usepackage{amsmath, amssymb}
\usepackage[dvipsnames]{xcolor,colortbl}
\usepackage{subfigure}
\graphicspath{{figures/}}
\newcommand{\orcid}[1]{\href{https://orcid.org/#1}{\includegraphics[width=10pt]{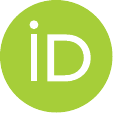}}}

\usepackage{xcolor}\colorlet{linkequation}{blue}
\usepackage[colorlinks=true, 
linkcolor=red,   
filecolor=red,   
citecolor=green, 
urlcolor=blue,
hyperfootnotes=true]{hyperref}

\newcommand*{\SavedEqref}{}
\let\SavedEqref\eqref
\renewcommand*{\eqref}[1]{%
	\begingroup
	\hypersetup{
		linkcolor=linkequation,
		linkbordercolor=linkequation,
	}%
	\SavedEqref{#1}%
	\endgroup
}

\begin{document}
\title{Exploring Anomaly-Free Dark Photon Models Through Solar Neutrino-Electron Scattering in Ton-Scale Dark Matter Detectors}
\newcommand{\ktu}{Department of Physics,
Karadeniz Technical University, Trabzon, TR61080, Türkiye}
\newcommand{\ktufb}{Graduate School of Natural and Applied Science, Karadeniz Technical University, Trabzon, TR61080, Türkiye}
\newcommand{\cern}{Theoretical Physics Department, CERN, 1211 Geneva 23, Switzerland}
\author{M.~Demirci \orcid{0000-0003-2504-6251}} %
\email{mehmetdemirci@ktu.edu.tr (corresponding author)}
\affiliation{\ktu}
\affiliation{\cern}
\author{G.~Kazıl}
\email{gurselkazil@ktu.edu.tr}
\affiliation{\ktu}
\affiliation{\ktufb}

\author{M.~F.~Mustamin}
\email{mfmustamin@ktu.edu.tr}
\affiliation{\ktu}

\date{\today}
\begin{abstract}
We study the phenomenological implications of anomaly-free dark photon models using low-energy solar neutrino-electron elastic scattering data from the ton-scale PandaX-4T and XENONnT experiments. Focusing on $U(1)_X$ gauge extensions that incorporate both kinetic and mass mixing, we derive novel exclusion limits across the parameter spaces of six benchmark models. Our analysis demonstrates that the stringency of these limits is directly correlated with the theoretical leptonic charge assignments of the respective models. Additionally, we provide model-independent limits by isolating the pure kinetic and pure mass mixing scenarios.  In the sub-MeV mediator mass regime, our results meaningfully improve upon existing limits from previous neutrino-electron scattering experiments, such as CHARM-II, TEXONO, and GEMMA. Ultimately, these findings yield improved limits on low-mass dark photon models, demonstrating their complementarity with other ventures in the global search for dark portals.
\end{abstract}


\maketitle
\section{Introduction}
The Standard Model (SM) of particle physics provides an elegant and remarkably successful description of the electroweak and strong interactions. Nevertheless, several outstanding observational facts---most notably the evidence of dark matter (DM), dark energy, and the baryon asymmetry of the universe---call for an extension beyond the SM (BSM). A neutral gauge boson, generically denoted as $Z'$, is one of the most commonly invoked new particles in such extensions. It naturally arises whenever the SM gauge group is enlarged by an extra Abelian factor or by a non-Abelian group that contains one~\citep{Langacker:2008yv}. 

Closely related to this concept is the so-called \textit{dark photon} \citep{Fabbrichesi:2020}. This particle was originally introduced as a light vector boson that couples to the SM exclusively through a small kinetic mixing with the QED field strength tensor~\citep{Holdom1986}. If kinetic mixing were the only new ingredient, the physical dark photon would inherit purely vectorial couplings to the electrically charged SM fermions and would not couple to SM neutrinos at tree level~\cite{Pospelov:2008zw}. However, in realistic ultraviolet (UV) completions, the dark photon is often associated with a spontaneously broken $U(1)_X$ gauge symmetry under which some SM fields may carry non-trivial charges. In such scenarios, a mass-mixing term between the SM $Z$ boson and the new gauge boson is generally present in addition to the kinetic mixing. This brings the setup back to the generic $Z'$ class and, crucially, induces couplings between the new mediator and SM neutrinos~\cite{Foot:1994vd, delAguila:1995rb, Babu:1997st}. Consequently, comprehensive phenomenological studies of generic $Z'$ or dark photon models must account for the simultaneous presence of both kinetic and mass mixing between the neutral vector bosons. Throughout this work, we therefore treat the $Z'$ and the dark photon interchangeably within a unified $U(1)_X$ framework that incorporates both of these mixing effects. While searching for this particle has been the subject of extensive efforts at colliders \citep{Chen:2020bok,Asai:2022zxw,Asai:2023mzl}, nuclear reactors \citep{Bilmis:2015lja,Lindner:2018kjo,Coloma:2022avw}, and spallation neutron sources \citep{AtzoriCorona:2022moj, Elpe:2022hqp, DeRomeri:2022twg, AtzoriCorona:2025ibl}, our focus lies on scrutinizing its parameter space using recent results from DM searches.

DM direct detection (DD) experiments are currently undergoing rapid advancements, significantly pushing the boundaries of rare-event searches. In these experiments, solar neutrinos constitute an irreducible background flux that induces elastic neutrino-electron scattering (E$\nu$ES) \citep{tHooft:1971ucy} signals within the low-energy electron recoil region~\citep{Baudis:2013qla,Billard:2013qya}. In this specific energy window, the solar neutrino flux is dominated by the high-rate pp chain, with sub-leading yet non-negligible contributions from the $^7\mathrm{Be}$ and $^8\mathrm{B}$ channels, as predicted to percent-level precision by modern standard solar models~\citep{Bahcall:1989ks, Bahcall:1995mm, Bahcall:2000nu, Bahcall:2004mq, Bahcall:2004pz, Vinyoles:2016djt}. The PandaX \citep{PandaX:2014mem, PandaX-II:2017hlx, PandaX:2018wtu, PandaX:2022ood, PandaX:2024cic, PandaX:2024med} and XENON \citep{XENON:2007uwm, XENON:2015gkh, XENON:2017vdw, XENON:2019gfn, XENON:2020rca, XENON:2022ltv} collaborations stand at the frontier of this effort, employing multi-tonne liquid xenon (LXe) time projection chambers (TPCs) characterized by massive exposures and ultra-low backgrounds in the sub-keV regime. Although DD facilities were originally optimized to search for dark matter candidates via nuclear recoils, the unprecedented suppression of their electron recoil backgrounds has transformed them into highly competitive observatories for new physics signatures. Recent studies have actively exploited this pristine low-background environment to constrain non-standard neutrino interactions \citep{Li:2024iij, Amaral:2023tbs, Celestino-Ramirez:2025snn}, light novel mediators \citep{Harnik:2012ni, Cerdeno:2016sfi, Demirci:2023tui, DeRomeri:2024dbv, Demirci:2025qdp}, and neutrino electromagnetic properties \citep{Khan:2023b, Giunti:2023yha, Demirci:2025poc}.

Motivated by the phenomenological relevance of the dark-photon/$Z'$ class of models and the recent availability of high-quality low-energy data, we investigate the signatures of dark photon models via the solar neutrino-induced E$\nu$ES process. Accordingly, we utilize the novel electronic-recoil datasets from PandaX-4T (Run0 and Run1)~\citep{PandaX:2024cic} and XENONnT (SR0)~\citep{XENON:2022ltv}. We consider both the kinetic-mixing and the general $U(1)_X$ scenario featuring simultaneous mass mixing. To connect with explicit UV-complete realizations, we adopt the family of anomaly-free $U(1)_X$ charge assignments constructed in Ref.~\citep{Campos:2017dgc} as benchmarks for dark photon models. 
These models were originally derived from gauge principles within two-Higgs-doublet models (2HDMs) to simultaneously generate non-zero neutrino masses through the seesaw mechanism and forbid tree-level flavor-changing neutral currents. These charge assignments---which include $U(1)_{B-L}$ as a particular case---are representative of a broad class of $U(1)_X$ extensions of the SM and were notably used as reference models in the related E$\nu$ES analysis \citep{Lindner:2018kjo}, as well as its implementation in coherent elastic-neutrino nucleus scattering \citep{Elpe:2022hqp}. For broader perspectives on dark photon models across terrestrial, astrophysical, and cosmological scales, we refer the reader to the extensive reviews in Refs.~\citep{Ilten:2018crw, Bauer:2018onh, Redi:2022zkt, Caputo:2025avc}.

The remainder of this paper is organized as follows. We introduce the theoretical formulation in Sec.~\ref{sec:nue} and describe the statistical methodology along with the experimental datasets in Sec.~\ref{sec:stat}. We then present our results in Sec.~\ref{sec:resdis}, discussing the expected event spectra, the derived upper limits on the allowed parameter space, and a comparison with previous bounds. Finally, we conclude our findings in Sec.~\ref{sec:summ}.

\section{General Formalism}\label{sec:nue}
\subsection{Dark Photon Models}
We focus on a class of light dark photon models in which the dark sector interacts with the SM via a vector portal. To realize this framework, we consider a general $U(1)$ extension of the SM based on the expanded gauge group $SU(2)_L \times U(1)_Y \times U(1)_X$. In the minimal setup, this portal interaction arises naturally through a gauge-invariant, renormalizable kinetic and/or mass mixing term between the gauge fields of the SM hypercharge $U(1)_Y$ and the new $U(1)_X$ symmetry~\cite{Holdom1986}. The new gauge boson associated with the $U(1)_X$ symmetry is naturally identified as the dark photon. The relevant Lagrangian in the gauge basis can be expressed as
\begin{align}
\begin{split}
\mathcal{L}\supset& - \frac{1}{4}\hat{W}_{\mu\nu}\hat{W}^{\mu\nu}-\frac{1}{4}\hat{B}_{\mu\nu} \hat{B}^{\mu\nu} - \frac{1}{4}\hat{X}_{\mu\nu}\hat{X}^{\mu\nu} \\ &- \frac{\epsilon}{2}\hat{B}_{\mu\nu}\hat{X}^{\mu\nu} + \sum_f \overline{f} i\gamma^\mu D_\mu f + \sum_\phi |D_\mu \phi|^2
\end{split}
\end{align}
where the gauge fields corresponding to the $SU(2)_L$, $U(1)_Y$, and $U(1)_X$ groups are denoted by $\hat{W}$, $\hat{B}$, and $\hat{X}$, respectively. The chiral fermion multiplets of the model are collectively represented by $f$ [e.g., $(\nu_L, e_L)^T, \nu_R, e_R$], and the scalar sector is denoted by $\phi$. The associated covariant derivative is given by
\begin{equation}
D_\mu = \partial_\mu + ig\sum_{a=1}^{3} \frac{\sigma^a}{2} \hat{W}_\mu^a + ig'\frac{Q_Y}{2} \hat{B}_\mu + i g_X \frac{Q_X}{2} \hat{X}_\mu \,,
\end{equation}
where $\sigma^a$ are the Pauli matrices corresponding to the $SU(2)_L$ generators. The parameters $g$, $g'$, and $g_X$ denote the gauge couplings for the $SU(2)_L$, $U(1)_Y$, and $U(1)_X$ groups, respectively, while $Q_Y$ and $Q_X$ specify the respective gauge charges of the interacting fields.
Notably, the two $U(1)$ field strength tensors, $\hat{B}_{\mu\nu}$ and $\hat{X}^{\mu\nu}$, couple to each other via the $-(\epsilon/2) \hat{B}_{\mu\nu}\hat{X}^{\mu\nu}$ term, which induces kinetic mixing between the new gauge boson and the SM hypercharge boson. The presence of this mixing mechanism is theoretically well-motivated; even if it is assumed to vanish at some higher energy scale, it is dynamically generated at the loop level by any massive state charged under both $U(1)$ groups \citep{Holdom1986}.

The mass of the new gauge boson arises following the spontaneous breaking of the $U(1)_X$ symmetry. To properly describe the physical states, it is necessary to transition from the fundamental gauge basis---composed of the neutral fields $\hat{W}^3$, $\hat{B}$, and $\hat{X}$---to the physical mass basis, comprising the photon ($A$), the SM-like $Z$ boson, and the new dark photon ($Z'$). Upon this transformation, all gauge bosons acquire canonical kinetic terms and become mass eigenstates, with the mixing parameters being explicitly absorbed into the physical couplings. While the general formalism for neutral-boson mixing has been extensively discussed in the literature \citep{Foot:1994vd, Babu:1997st}, we adopt the exact analytical framework established in Ref.~\cite{Lindner:2018kjo} for a comprehensive derivation of this basis transformation.

In this physical basis, the relevant neutral-current interaction Lagrangian\footnote{Note that the charged-current interactions mediated by the $W^\pm$ bosons remain identical to those in the SM.} can be expressed compactly as
\begin{equation}
\mathcal{L}_{\text{NC}} \supset -J_\text{em}^\mu A_\mu - J_Z^\mu Z_\mu - J_{Z'}^\mu Z'_\mu \,,
\end{equation}
where $J_\text{em}^\mu$ is the standard electromagnetic current interacting with the photon field $A_\mu$. The massive eigenstates, $Z_\mu$ and $Z'_\mu$, couple to the fermions via the currents $J_Z^\mu$ and $J_{Z'}^\mu$, respectively. These physical currents are related to the SM neutral current $J_\text{NC}^\mu$ and the new $U(1)_X$ current $J_X^\mu$ through a rotation matrix:
\begin{align}
J_Z^\mu &= c_\alpha J_\text{NC}^\mu - s_\alpha J_X^\mu \,, \\
J_{Z'}^\mu &= s_\alpha J_\text{NC}^\mu + c_\alpha J_X^\mu \,,
\end{align}
where the shorthand notations $c_\alpha \equiv \cos\alpha$ and $s_\alpha \equiv \sin\alpha$ are used, with $\alpha$ representing the mass mixing angle between the SM $Z$ boson and the dark photon $Z'$. 
Explicitly, the explicit forms of the three currents are given by
\begin{align}
J_\text{em}^\mu &= g s_W \sum_f \bar{f} \gamma^\mu Q_\text{em}^f f \,, \\ 
J_\text{NC}^\mu &= g \sum_f \bar{f} \gamma^\mu \left( c_W Q_\text{em}^f - \frac{Q_Y^f}{2c_W}\right) f \,, \\
J_X^\mu &= \sum_f \bar{f} \gamma^\mu \frac{c_W g_X Q_X^f - g\epsilon Q_Y^f s_W}{2\sqrt{1-\epsilon^2} c_W} f \,,
\end{align}
where the weak mixing angle $\theta_W$ is defined via $s_W\equiv \sin\theta_W$ and $c_W\equiv \cos\theta_W$, which is related by $g'=(s_W/c_W) g$.
Furthermore, $Q_\text{em}^f$ denotes the electromagnetic charge, $Q_Y^f$ the weak hypercharge, and $Q_X^f$ the additional charge associated with the $U(1)_X$ gauge group.

The specific values of the $U(1)_X$ charges depend on the particular dark photon model under consideration and must be carefully assigned to ensure exact gauge anomaly cancellation. To this end, we adopt the anomaly-free charge assignments originally formulated in the context of 2HDMs devoid of tree-level flavor-changing neutral currents~\citep{Campos:2017dgc}. We discuss details of this process in Appendix~\ref{app:anomaly_free}.
\begin{table}[htb]
	\centering
    \caption{Summary of the anomaly-free $U(1)_X$ gauge charge assignments for the benchmark dark photon models considered in this work. These specific theoretical configurations are adapted from Ref.~\cite{Campos:2017dgc}.}
	\label{tab:dp_charges}
	\footnotesize 
	\begin{ruledtabular}
	\begin{tabular}{lcccccc}
		\textbf{Charges}  & \boldmath{$U(1)_{B-L}$} & \boldmath{$U(1)_C$} & \boldmath{$U(1)_D$} & \boldmath{$U(1)_E$} & \boldmath{$U(1)_F$} & \boldmath{$U(1)_G$} \\
		\midrule   
		$Q_{X\ell}^L$  & $-1$ & $\frac{3}{4}$ & $-\frac{3}{2}$ & $-\frac{3}{2}$ & $-3$ & $-\frac{1}{2}$ \\[1ex]
		$Q_{X e}^R$ & $-1$ & $0$   & $-2$    & $-1$   & $-4$ & $0$ \\
	\end{tabular}
	\end{ruledtabular}
\end{table}
The specific $U(1)_X$ gauge charge assignments for each benchmark model are summarized in Table~\ref{tab:dp_charges}. In this notation, the index $X$ specifies the particular dark photon model under consideration, while the parameters $Q_{X\ell}^L$ and $Q_{Xe}^R$ denote the gauge charges associated with the left-handed lepton doublet, $(\nu_L, e_L)^T$, and the right-handed electron singlet, $e_R$, respectively.
Crucially, these specific anomaly-free configurations were constructed not only to resolve the flavor problem within the 2HDM framework---by naturally forbidding tree-level flavor-changing neutral currents---but also to successfully accommodate non-zero neutrino masses via the seesaw mechanism.

To facilitate our analysis, we explicitly cast the relevant interaction Lagrangian for the massive vector bosons into a chiral basis. By decomposing the fermion fields into their left- and right-handed components, $f_{L,R} = P_{L,R} f$, with the projection operators $P_{L,R} = (1 \mp \gamma^5)/2$, the interaction Lagrangian involving the $W^\pm$, $Z$, and $Z'$ bosons can be expressed as
\begin{align}
\begin{split}
\mathcal{L} \supset &- \Big( W^+_\mu \bar{\nu} \gamma^\mu P_L C_{\nu \ell W}^L \ell + \text{h.c.} \Big) \\
&- \Big[ \bar{\nu} \gamma^\mu P_L C_{\nu \nu Z}^L \nu + \bar{\ell} \gamma^\mu\left(C^L_{\bar{\ell} \ell Z} P_L + C^R_{\bar{\ell} \ell Z} P_R\right) \ell \Big] Z_\mu \\ 
&- \Big[ \bar{\nu} \gamma^\mu P_L C_{\nu \nu Z'}^L \nu + \bar{\ell} \gamma^\mu\left(C^L_{\bar{\ell} \ell Z'}P_L + C^R_{\bar{\ell} \ell Z'} P_R\right) \ell \Big] Z'_\mu \,.
\end{split}
\label{eq:dp_lagrangian}
\end{align}
Here, we have defined the effective chiral couplings governing the leptonic interactions with the gauge bosons as
\begin{align}
C_{\nu \ell W}^L &= \frac{g}{\sqrt{2}} \,, \\
C_{\nu \nu Z}^L &= \frac{gc_\alpha}{2c_W} - \frac{g\epsilon s_W s_\alpha}{2c_W \sqrt{1-\epsilon^2}} - \frac{g_X s_\alpha Q_{X\nu}^L}{2\sqrt{1-\epsilon^2}} \,, \\
C_{\nu \nu Z'}^L &= \frac{gs_\alpha}{2c_W} + \frac{g\epsilon s_W c_\alpha}{2c_W\sqrt{1-\epsilon^2}} + \frac{g_X c_\alpha Q_{X\nu}^L}{2\sqrt{1-\epsilon^2}} \,.
\end{align}
Similarly, the left- and right-handed couplings for the charged lepton $\ell$ (which strictly corresponds to the electron in the context of our scattering process) interacting with the SM-like $Z_\mu$ boson are parameterized as
\begin{align}
C^L_{\ell\ell Z} &= \frac{gc_\alpha(2s_W^2-1)}{2c_W} - \frac{g\epsilon s_W s_\alpha}{2c_W \sqrt{1-\epsilon^2}} - \frac{g_X s_\alpha Q_{X\ell}^L}{2\sqrt{1-\epsilon^2}} \,, \\
C^R_{\ell\ell Z} &= \frac{gc_\alpha s_W^2}{c_W} - \frac{g\epsilon s_W s_\alpha}{c_W \sqrt{1-\epsilon^2}} - \frac{g_X s_\alpha Q_{X\ell}^R}{2\sqrt{1-\epsilon^2}} \,,
\end{align}
while the analogous chiral couplings for the new dark gauge boson $Z'_\mu$ read
\begin{align}
C^L_{\ell\ell Z'} &= \frac{gs_\alpha (2s_W^2-1)}{2c_W} + \frac{g\epsilon s_W c_\alpha}{2c_W\sqrt{1-\epsilon^2}} + \frac{g_X c_\alpha Q_{X\ell}^L}{2\sqrt{1-\epsilon^2}} \,, \\
C^R_{\ell\ell Z'} &= \frac{gs_\alpha s_W^2}{c_W} + \frac{g\epsilon s_W c_\alpha}{c_W \sqrt{1-\epsilon^2}} + \frac{g_X c_\alpha Q_{X\ell}^R}{2\sqrt{1-\epsilon^2}} \,.
\end{align}

\subsection{Elastic Neutrino-Electron Scattering}

The $\text{E}\nu\text{ES}$ process offers a theoretically clean, purely leptonic window into electroweak precision measurements~\citep{Formaggio:2012cpf}. Among dedicated terrestrial measurements, the most precise determination of $\nu_\mu$ ($\bar{\nu}_\mu$)--electron elastic scattering originates from the CHARM-II experiment, which achieved a precision of approximately $3\%$ on the electroweak couplings~\cite{CHARM-II:1993phx, CHARM-II:1994dzw}. Meanwhile, the most rigorous measurements of $\nu_e$--$e$ and $\bar{\nu}_e$--$e$ scattering are provided by the LSND~\cite{LSND:2001akn} and TEXONO~\cite{TEXONO:2009knm} experiments, respectively, both operating at the $\sim 20\%$ precision level. Beyond its importance for precision electroweak tests, the $\text{E}\nu\text{ES}$ process is physically distinguished by its highly directional kinematics. The scattered electron emerges at a very narrow forward angle relative to the incoming neutrino trajectory---a distinct topological feature that has been extensively exploited in large-scale solar neutrino experiments, including Super-Kamiokande~\cite{Kamiokande-II:1989hkh, Super-Kamiokande:2001ljr}, SNO~\cite{SNO:2002tuh, SNO:2003bmh}, and BOREXINO~\cite{Borexino:2007kvk, Borexino:2017rsf}.

Capitalizing on its theoretically clean nature, the $\text{E}\nu\text{ES}$ process provides an excellent avenue to probe beyond the SM physics, such as dark photon models. It should be emphasized that neither kinetic nor mass mixing alters the SM charged-current (CC) channel; hence, the $W$-boson exchange remains standard. However, in the neutral-current (NC) sector, neutrinos can additionally interact with electrons via the $Z$ and $Z'$ gauge bosons (where we generically refer to the dark photon-like mediators as $Z'$ bosons). To formally evaluate these BSM contributions, we must determine the exact scattering amplitudes. Based on the Lagrangian provided in Eq.~\eqref{eq:dp_lagrangian}, the scattering amplitudes for $\text{E}\nu\text{ES}$ mediated by the $W$, $Z$, and $Z'$ bosons in the low-energy limit can be expressed as
\begin{align}
i\mathcal{M}_W &= \left[\bar{u}_\nu \left(i\frac{g}{\sqrt{2}}\gamma^\mu P_L \right) u_\nu \right] \left[ \frac{ig_{\mu\rho}}{m_W^2}\right]\left[ \bar{u}_e \left(i\frac{g}{\sqrt{2}}\gamma^\rho P_L\right)u_e\right] \,, \\
i\mathcal{M}_Z &= \left[\bar{u}_\nu \big(i \gamma^\mu P_L C_{\nu \nu Z}^L\big) u_\nu\right] \left[\frac{ig_{\mu\rho}}{m_Z^2}\right] \nonumber \\
&\quad \times \left[\bar{u}_e \big(i \gamma^\rho (C^L_{\ell\ell Z} P_L + C^R_{\ell\ell Z} P_R) \big) u_e\right] \,, \\
i\mathcal{M}_{Z'} &= \left[\bar{u}_\nu \big(i \gamma^\mu P_L C_{\nu \nu Z'}^L\big) u_\nu\right] \left[\frac{ig_{\mu\rho}}{2m_e T_e + m_{Z'}^2}\right] \nonumber \\
&\quad \times \left[ \bar{u}_e \big(i \gamma^\rho (C^L_{\ell\ell Z'}P_L + C^R_{\ell\ell Z'} P_R) \big) u_e\right] \,. \label{eq:MZprime}
\end{align}
For the $W$-boson contribution, we have applied a Fierz transformation to ensure structural consistency with the neutral-current terms arising from the other gauge bosons. By combining these individual scattering amplitudes and utilizing the standard Fermi constant relation, $G_F/\sqrt{2} = g^2/(8m_W^2)$, the total scattering amplitude can be compactly cast as
\begin{equation}
	\mathcal{M}_\text{total} = -\frac{4G_F}{\sqrt{2}}\left[\bar{u}_\nu \gamma^\mu P_L u_\nu \right] \left[\bar{u}_e \gamma_\mu \left(\widetilde{g}_L P_L + \widetilde{g}_R P_R\right) u_e\right] \,.
\end{equation}
Here, we have defined the generalized effective chiral couplings, $\widetilde{g}_L$ and $\widetilde{g}_R$, as
\begin{align} 
\widetilde{g}_L &= \delta_{\ell e} + \frac{2c_W^2}{g^2} C_{\nu \nu Z}^L C^L_{\ell\ell Z} + \frac{\sqrt{2} C_{\nu \nu Z'}^L C^L_{\ell\ell Z'} }{4G_F (2m_e T_e + m_{Z'}^2)} \,, \\
\widetilde{g}_R &= \frac{2c_W^2}{g^2} C_{\nu \nu Z}^L C^R_{\ell\ell Z} + \frac{\sqrt{2} C_{\nu \nu Z'}^L C^R_{\ell\ell Z'} }{4G_F (2m_e T_e + m_{Z'}^2)} \,.
\end{align}
Because this total amplitude exhibits the exact same chiral structure as the standard electroweak amplitude, the resulting differential cross section---obtained by squaring the matrix element and summing over the spin states---retains the familiar SM functional form. 
Consequently, the differential cross section with respect to the electron recoil energy $T_e$ is given by
\begin{equation} 
	\frac{d\sigma_{\nu_\ell}}{dT_e} = \frac{2G_F^2 m_e}{\pi} \left[ \widetilde{g}_L^2 + \widetilde{g}_R^2 \left(1-\frac{T_e}{E_\nu}\right)^2 - \widetilde{g}_L \widetilde{g}_R \frac{m_e T_e}{E_\nu^2} \right] \,,
	\label{eq:sm_nue}
\end{equation}
where $m_e$ denotes the electron mass and $E_\nu$ is the incoming neutrino energy. We can explicitly expand the generalized effective chiral couplings, $\widetilde{g}_{L}$ and $\widetilde{g}_{R}$, in terms of the underlying model parameters as follows:
\begin{widetext}
	\begin{align}
		\begin{split}
			\widetilde{g}_L = \delta_{\ell e} &+ (s_W^2-\frac{1}{2})c^2_\alpha -\frac{c_\alpha s_\alpha}{\sqrt{1-\epsilon^2}}\left[ \epsilon s_W^3 + \frac{c_W g_X}{2g} (Q_{X\ell}^L + (2s_W^2-1) Q_{X\nu}^L) \right] \\ & + \frac{s_\alpha^2}{2(1-\epsilon^2)}\left[\epsilon^2 s_W^2 + \frac{2c_W s_W \epsilon g_X}{g} (Q_{X\nu}^L+Q_{X\ell}^L) + \frac{c_W^2 g_X^2}{g^2} Q_{X\nu}^L Q_{X\ell}^L\right] \\
			&+ \frac{1}{2\sqrt{2}G_F (2m_e T_e + m_{Z'}^2)} \Bigg\{s_\alpha^2 \frac{g^2(2s_W^2-1)}{4c_W^2} + \frac{c_\alpha s_\alpha}{\sqrt{1-\epsilon^2}}\left[\frac{ g^2 s_W^3 \epsilon}{2c_W^2} + \frac{g g_X}{4c_W}(Q_{X\ell}^L + (2s_W^2-1)Q_{X\nu}^L) \right] \\
			&+ \frac{c_\alpha^2}{1-\epsilon^2}\left[\frac{g^2s_W^2\epsilon^2}{4c_W^2} + \frac{g s_W g_X \epsilon}{4c_W}(Q_{X\nu}^L+Q_{X\ell}^L) + \frac{g_X^2}{4} Q_{X\nu}^L Q_{X\ell}^L \right] \Bigg\}
		\end{split}
		\label{eq:gL_SMDP}
	\end{align}
	\begin{align}
		\begin{split}
			\widetilde{g}_R =  & s_W^2 c_\alpha^2 -\frac{c_\alpha s_\alpha}{\sqrt{1-\epsilon^2}} \left[ \epsilon s_W(1+s_W^3) + \frac{c_W g_X}{2g} (Q_{X\ell}^R + 2s_W^2 Q_{X\nu}^L) \right] \\ 
			& + \frac{s_\alpha^2}{2(1-\epsilon^2)} \left[\epsilon^2 s_W^2 + \frac{c_W s_W \epsilon g_X}{g} (2Q_{X\nu}^L+Q_{X\ell}^R) + \frac{c_W^2 g_X^2}{g^2} Q_{X\nu}^L Q_{X\ell}^R\right] \\
			&+ \frac{1}{2\sqrt{2}G_F (2m_e T_e + m_{Z'}^2)} \Bigg\{s_\alpha^2 \frac{g^2s_W^2}{2c_W^2} + \frac{c_\alpha s_\alpha}{\sqrt{1-\epsilon^2}}\left[\frac{ g^2 s_W \epsilon (1+s_W^2)}{2c_W^2} + \frac{g g_X}{4c_W}(Q_{X\ell}^R + 2s_W^2 Q_{X\nu}^L) \right] \\
			&+ \frac{c_\alpha^2}{1-\epsilon^2}\left[\frac{g^2 s_W^2\epsilon^2}{2c_W^2} + \frac{g s_W g_X \epsilon}{4c_W}(2Q_{X\nu}^L+Q_{X\ell}^R) + \frac{g_X^2}{4} Q_{X\nu}^L Q_{X\ell}^R \right] \Bigg\}.
		\end{split}
		\label{eq:gR_SMDP}
	\end{align}
\end{widetext}
Both effective couplings depend on the flavor $\ell = e, \mu, \tau$ of the incoming neutrino. These expressions explicitly encode the individual contributions from both the SM and the dark photon model, alongside the resulting interference terms. In the above relations, the Kronecker delta symbol $\delta_{\ell e}$ elegantly accounts for the distinct CC and NC channels of the process. Specifically, while both channels contribute to scattering processes involving $\nu_e$, only the NC channel is present for the $\nu_\mu$ and $\nu_\tau$ flavors. For our numerical evaluations, we adopt the weak mixing angle $s^2_W = 0.23857$ defined in the $\overline{\text{MS}}$ scheme~\cite{ParticleDataGroup:2024cfk}. 

It is worth noting that the general dark photon contributions to the $\text{E}\nu\text{ES}$ process gracefully simplify under specific limits:
\begin{itemize}
	\item For $\alpha=\epsilon=0$, the framework reduces to the SM augmented by a universal $Z'$.
	\item For $\alpha=\epsilon=g_X=0$, one exactly recovers the pure SM scenario, which correctly yields
	\begin{equation}
		\widetilde{g}_L \to \delta_{\ell e} + \left(s_W^2 - \frac{1}{2}\right) \quad \text{and} \quad \widetilde{g}_R \to s_W^2 \,.
	\end{equation}
\end{itemize}

Furthermore, we emphasize that our analytical results are in full agreement with those reported in Ref.~\cite{Lindner:2018kjo}, thereby validating our chiral-basis formulation.
As evident from the propagator in Eq.~\eqref{eq:MZprime}, the amplitude's dependence on $m_{Z'}$ is strictly governed by the $2m_e T_e + m_{Z'}^2$ term. Consequently, in the kinematic regime where $2m_e T_e \gg m_{Z'}^2$, the scattering process becomes entirely insensitive to the dark photon mass. This implies that a direct detection experiment with a recoil energy threshold $T_\text{min}$ effectively loses mass sensitivity below $m_{Z'}^\text{min} \simeq \sqrt{2 m_e T_\text{min}}$, rendering any observable new physics effects mass-independent in this sub-MeV region.

In DD experiments designed for dark matter searches, incoming solar neutrinos inevitably interact with the atomic electrons of the detector material, thereby triggering E$\nu$ES events. This process constitutes a prominent component of the pervasive neutrino background, often competing with potential DM recoil signals. Consequently, a precise understanding and characterization of these $\text{E}\nu\text{ES}$ events are crucial for robustly isolating authentic DM signatures. Furthermore, because the SM prediction for this scattering process is known with exceptional accuracy, any measured deviation in the electron recoil spectrum serves as a direct and highly sensitive probe for BSM physics, particularly for the $U(1)_X$ dark photon models investigated in this work.

\section{Data Analysis Details}\label{sec:stat}

In this section, we outline the statistical framework employed to derive limits from the low-energy electronic recoil datasets of the PandaX-4T~\cite{PandaX:2024cic} and XENONnT~\citep{XENON:2022ltv} experiments. These observed recoil events are primarily driven by the irreducible solar neutrino flux. 

Located at the China Jinping Underground Laboratory, the PandaX-4T experiment utilizes a dual-phase time projection chamber (TPC) containing $3.7$ tonnes of active liquid xenon (LXe) in its sensitive volume~\cite{PandaX:2018wtu}. The collaboration recently released two distinct datasets covering electron recoil energies up to $30$~keV. The first dataset, denoted as Run0, stems from the commissioning phase, operating from November 28, 2020, to April 16, 2021. The subsequent dataset, Run1, spans the primary data-taking period from November 16, 2021, to May 15, 2022. Similarly, the XENONnT experiment, situated at the Laboratori Nazionali del Gran Sasso, employs comparable LXe TPC technology with a significantly larger active mass of $5.9$ tonnes~\citep{XENON:2022ltv}. The XENON collaboration reported the SR0 dataset, which targets a continuous energy range up to $30$~keV, collected between July 6, 2021, and November 10, 2021. Since their inception, both state-of-the-art experiments have actively conducted a broad array of new physics searches, placing some of the most stringent constraints to date on weakly interacting massive particles, axion-like particles, neutrino electromagnetic properties, and dark photons.

Within this experimental context, we investigate the potential contributions of the anomaly-free $U(1)_X$ dark photon models to the expected $\text{E}\nu\text{ES}$ rate. These beyond the SM (BSM) effects are systematically analyzed across the low-energy electron recoil region of interest (ROI) specifically defined for each experiment. In what follows, we detail the computation of the theoretical recoil rates and delineate our statistical methodology.

\subsection{Recoil Rate Calculations}
The expected differential event rate for the $\text{E}\nu\text{ES}$ process, obtained by folding the differential scattering cross section with the incoming solar neutrino flux, can be expressed as
\begin{align}
	\left[\frac{dR}{dT_{e}}\right]^i_X = Z_{\text{eff}}(T_e) \int_{E_{\nu}^{\text{min}}}^{E_{\nu,i}^{\text{max}}} dE_\nu \frac{d\Phi^i}{dE_\nu} \left[\frac{d\sigma}{dT_{e}}\right]^{\nu e}_X.
\end{align}
Here, the minimum neutrino energy required to produce an electron recoil of energy $T_e$ is kinematically determined by
\begin{equation}
	E_{\nu}^{\text{min}} = \frac{1}{2}\left(T_e + \sqrt{T_e^2 + 2m_e T_e} \right) \,,
\end{equation}
while $E_{\nu,i}^{\text{max}}$ denotes the kinematic endpoint of the respective neutrino flux. The differential solar neutrino flux is represented by $d\Phi^i/dE_\nu$, where the index $i \in \{pp, ^7\text{Be}, \dots\}$ labels the distinct solar neutrino production channels. Given their dominant contribution to low-energy neutrino-electron scattering, we restrict our analysis in this work to the $pp$ and $^7\text{Be}$ solar neutrino fluxes. Additionally, the index $X \in \{B-L, C, D, \dots\}$ specifies the benchmark dark photon model under consideration.

Because the region of interest involves low electron recoil energies ($T_e \sim \text{a few keV}$) that are comparable to the atomic binding energies of the target material, the standard free-electron approximation is no longer strictly valid due to atomic ionization effects \cite{Chen:2016eab}. To properly account for this, the scattering cross section must be scaled by the effective number of atomic electrons available for ionization at a given recoil energy, defined as
\begin{equation}
	Z_{\text{eff}}(T_e) = \sum_{i} n_i \theta(T_e - B_i) \,.
\end{equation}
In this expression, $n_i$ and $B_i$ represent the number of electrons and their corresponding binding energy in the atomic shell, respectively, while $\theta(x)$ is the Heaviside step function. For our numerical evaluations, we adopt the shell structure and binding energy values for the liquid xenon target from Ref.~\citep{xraydata:2009}. Implementing this effective charge correction is crucial for appropriately modifying the scattering cross sections derived under the free-electron hypothesis, which inherently assumes that the target electrons are completely unbound and initially at rest \cite{Kouzakov:2017hbc, Hsieh:2019hug}.

During their propagation to Earth, solar neutrinos undergo flavor oscillations, arriving at the detector as an incoherent mixture of $\nu_e$, $\nu_\mu$, and $\nu_\tau$. To appropriately account for this phenomenon, the theoretical differential cross section must be weighted by the relevant oscillation probabilities:
\begin{align}
	\left[\frac{d\sigma}{dT_{e}}\right]^{\nu e}_X &= P_{ee} \left[\frac{d\sigma_{\nu_e}}{dT_{e}}\right]_X + \sum_{f=\mu,\tau} P_{ef} \left[\frac{d\sigma_{\nu_f}}{dT_{e}}\right]_X,
\end{align}
where $P_{e\mu} = (1-P_{ee})\cos^2\vartheta_{23}$ and $P_{e\tau} = (1-P_{ee})\sin^2\vartheta_{23}$. Here, $P_{ee}$ denotes the survival probability of the electron neutrino ($\nu_e$) at Earth, which is analytically given by
\begin{align}
	P_{ee} &= c_{13}^2 {c_{13}^m}^2\left( \frac{1}{2} + \frac{1}{2} \cos 2\vartheta_{12}^m \cos 2\vartheta_{12} \right) + s_{13}^2 {s_{13}^m}^2.
	\label{eq:Pee}
\end{align}
In this expression, $c_{13} \equiv \cos \vartheta_{13}$ and $s_{13} \equiv \sin \vartheta_{13}$, while the superscript $m$ indicates parameters modified by the solar matter effect (the Mikheyev-Smirnov-Wolfenstein mechanism). Specifically, $\vartheta_{12}^m$ represents the effective mixing angle in matter. In evaluating these survival probabilities, we also incorporate the day-night asymmetry arising from the Earth matter effect. The standard neutrino oscillation parameters are adopted from the global $3\nu$ analysis by NuFIT~6.0 \cite{Esteban:2024eli}.

The predicted number of elastic neutrino-electron scattering events in the $k$-th reconstructed energy bin is evaluated as
\begin{align}
	\begin{split}
		R^k_X &= \varepsilon N_{t} \int_{T_e^k}^{T_e^{k+1}} dT_e \, \mathcal{A}(T_e) \\
		&\quad \times \int_{0}^{T_e^{'\text{max}}} dT_e' \, \mathcal{R}(T_e,T_e') \sum_{i=pp,^7\text{Be}} \left[\frac{dR}{dT'_{e}}\right]^i_X,
	\end{split}
\end{align}
where $\varepsilon$ is the experimental exposure and $N_{t}$ denotes the number of target nuclei per unit mass of the detector material. The integration limit $T_e^{'\text{max}} = 2E_\nu^2/(2E_\nu + m_e)$ is the maximum kinematically allowed true recoil energy. The variables $T_e$ and $T_e'$ correspond to the reconstructed and true electron recoil energies, respectively. The detector response is rigorously characterized by the energy-dependent detection efficiency $\mathcal{A}(T_e)$ and the normalized Gaussian smearing function $\mathcal{R}(T_e,T_e')$, which accounts for the finite energy resolution of the instrument. 

For the PandaX-4T analysis, we adopt the efficiency curve from Ref.~\cite{PandaX:2024cic} and the energy resolution from Ref.~\cite{PandaX:2022ood}. Similarly, for XENONnT, the corresponding efficiency and energy resolution functions are taken from Ref.~\cite{XENON:2022ltv} and Ref.~\cite{XENON:2020rca}, respectively. Finally, the expected rates are scaled by their specific exposures: the PandaX-4T Run0 and Run1 datasets feature exposures of $198.9~\text{ton}\cdot\text{day}$ and $363.3~\text{ton}\cdot\text{day}$, respectively, whereas the XENONnT SR0 dataset was collected with an exposure of $1.16~\text{ton}\cdot\text{year}$.

\subsection{Statistical Methodology}\label{sec:chi2}
To statistically evaluate the new physics parameters of interest, denoted by $\mathcal{S}$, we construct tailored $\chi^2$ test statistics appropriate for the statistical nature of each dataset \cite{Baker:1983tu, Fogli:2002pt}. For the XENONnT data, which is adequately described by Gaussian statistics in its energy bins, the $\chi^2$ function is defined as
\begin{align}
	\begin{split}
		\chi^2(\mathcal{S}) &= \min_{\boldsymbol{\alpha}, \boldsymbol{\beta}} \Biggl[ \sum_{k=1}^{30} \left( \frac{ R_{\text{exp}}^{k}(\mathcal{S}; \boldsymbol{\alpha}, \boldsymbol{\beta}) - R_{\text{obs}}^{k} }{\sigma^{k}}\right)^2 \\ 
		&\quad + \sum_{i}\left(\frac{\alpha_i}{\sigma_{\alpha_i}}\right)^2 + \sum_{j}\left(\frac{\beta_j}{\sigma_{\beta_j}}\right)^2 \Biggr].
	\end{split} \label{eq:chi2_gauss}
\end{align}
Conversely, for the PandaX-4T data, we employ a Poisson-based likelihood ratio (the Baker-Cousins $\chi^2$ function), which is more robust for low-count statistics:
\begin{align}
	\begin{split}
		\chi^2(\mathcal{S}) &= \min_{\boldsymbol{\alpha}, \boldsymbol{\beta}} \Biggl[ 2 \sum_{k=1}^{30} \Biggl( R_{\text{exp}}^{k}(\mathcal{S}; \boldsymbol{\alpha}, \boldsymbol{\beta}) - R_{\text{obs}}^{k} \\ 
		&\quad + R_{\text{obs}}^{k} \ln \left[ \frac{ R_{\text{obs}}^{k}}{R_{\text{exp}}^{k}(\mathcal{S}; \boldsymbol{\alpha}, \boldsymbol{\beta})} \right] \Biggr) \\
		&\quad + \sum_{i}\left(\frac{\alpha_i}{\sigma_{\alpha_i}}\right)^2 + \sum_{j}\left(\frac{\beta_j}{\sigma_{\beta_j}}\right)^2 \Biggr].
	\end{split} \label{eq:chi2_poisson}
\end{align}

In these expressions, $R_{\text{obs}}^k$ and $R_{\text{exp}}^k$ represent the observed and theoretically expected event rates in the $k$-th energy bin, respectively. The expected event rate incorporates combined contributions from the SM, the proposed new physics (NP) scenarios, and various established background sources. To rigorously account for systematic uncertainties, we introduce the pull variables (nuisance parameters) $\alpha_i$ and $\beta_j$, such that the expected rate is parameterized as
\begin{align}
	\begin{split}
		R_{\text{exp}}^k(\mathcal{S}; \boldsymbol{\alpha}, \boldsymbol{\beta}) &= \sum_i (1+\alpha_i) \left[ R^{k,i}_{\text{SM}} + R^{k,i}_{\text{NP}}(\mathcal{S}) \right] \\
		&\quad + \sum_j (1+\beta_j) R_{\text{Bkg},j}^k.
	\end{split} \label{eq:rexp}
\end{align}
Here, the nuisance parameters $\alpha_i$ and $\beta_j$ absorb the fractional uncertainties related to the $i$-th solar neutrino flux component and the $j$-th background normalization, respectively. 

The experimental statistical uncertainty in the $k$-th bin is denoted by $\sigma^k$. The standard deviations constraining the nuisance parameters, $\sigma_{\alpha_i}$ and $\sigma_{\beta_j}$, are adopted from Ref.~\cite{Vinyoles:2016djt} for the solar neutrino fluxes and Ref.~\cite{PandaX:2024cic} for the background models. Finally, the combined SM and NP event rates are evaluated using the standard solar neutrino energy spectra \cite{Bahcall:1989ks}, appropriately normalized according to the B16-GS98 standard solar model \cite{Vinyoles:2016djt}.

\section{Results and Discussion}\label{sec:resdis}
In this section, we present the main results derived from the statistical framework outlined above. First, we illustrate the expected event spectra as functions of the electron recoil energy. Following this, we derive and present the $90\%$ CL limits (for 2 dof) on the parameter space of the considered dark photon models. 
Finally, we contextualize our findings by comparing them with existing constraints from a diverse array of complementary probes. These include previous neutrino-electron scattering experiments (CHARM-II~\citep{CHARM-II:1993phx, CHARM-II:1994dzw}, TEXONO~\citep{Wong:2004ru, TEXONO:2009knm}, BOREXINO \citep{Borexino:2007kvk, Borexino:2017rsf}, and GEMMA~\citep{Beda:2009kx, Beda:2012zz}), beam dump experiments (E141~\citep{Riordan:1987aw, Bjorken:2009mm}, $\nu$-cal~\citep{Blumlein:2011mv, Blumlein:2013cua}, and NA64~\citep{Banerjee:2019pds, NA64:2022yly}), collider searches (BaBar~\citep{BaBar:2014zli, BaBar:2017tiz} and LHCb~\citep{LHCb:2017trq, LHCb:2019vmc}), anomalous magnetic moment measurements for the electron ($(g-2)_e$)~\citep{Parker:2018vye, Morel:2020dww} and muon ($(g-2)_\mu$)~\citep{Tsai:2019buq, Muong-2:2021ojo}, atomic spectra precision tests~\citep{Bartlett:1988yy, Jaeckel:2010xx}, and astrophysical limits stemming from dark photons produced as thermal radiation in the Sun~\citep{Li:2023vpv}.

\begin{figure*}[htb]
	\centering
	\includegraphics[scale=0.38]{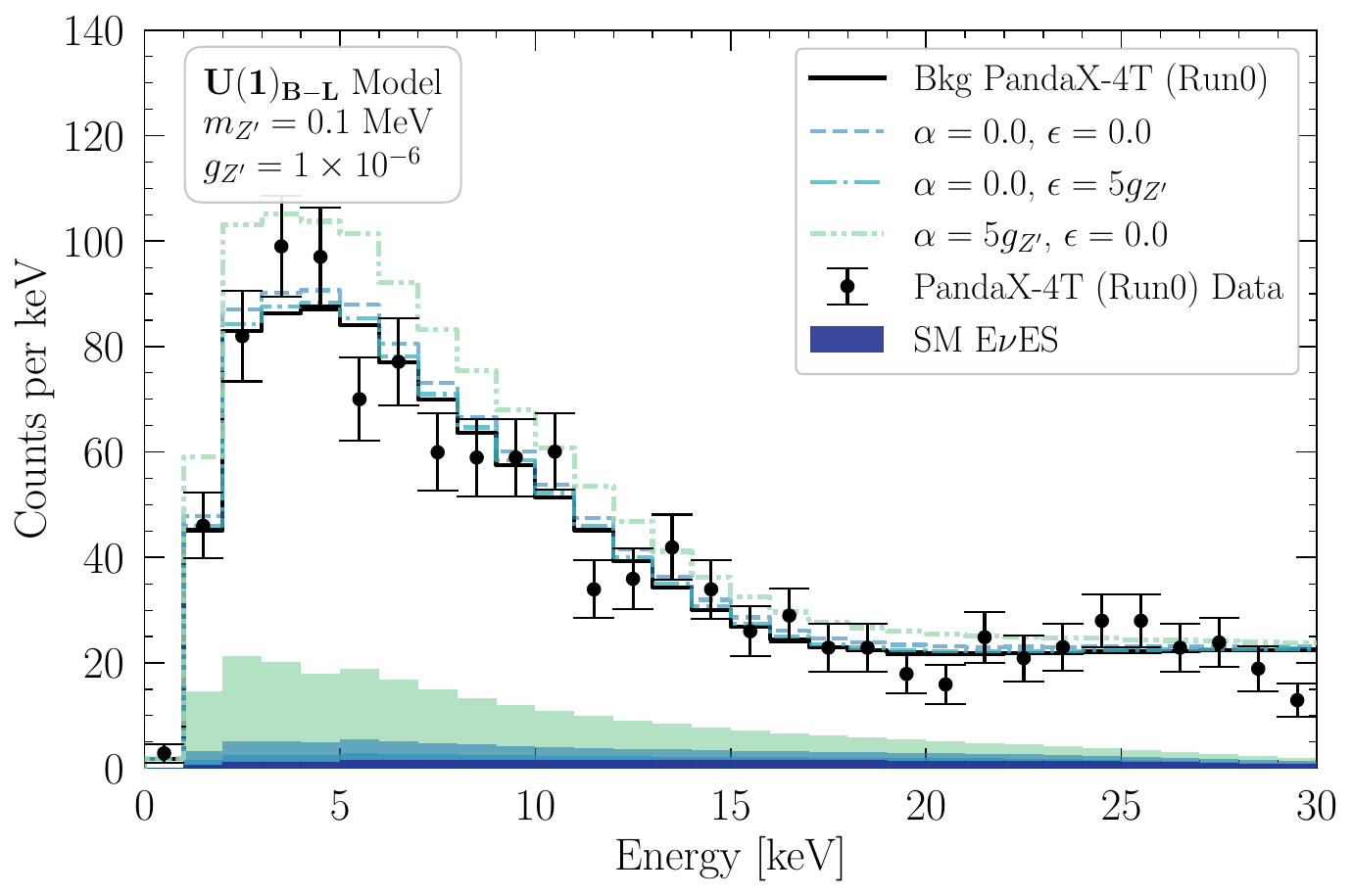}
	\includegraphics[scale=0.38]{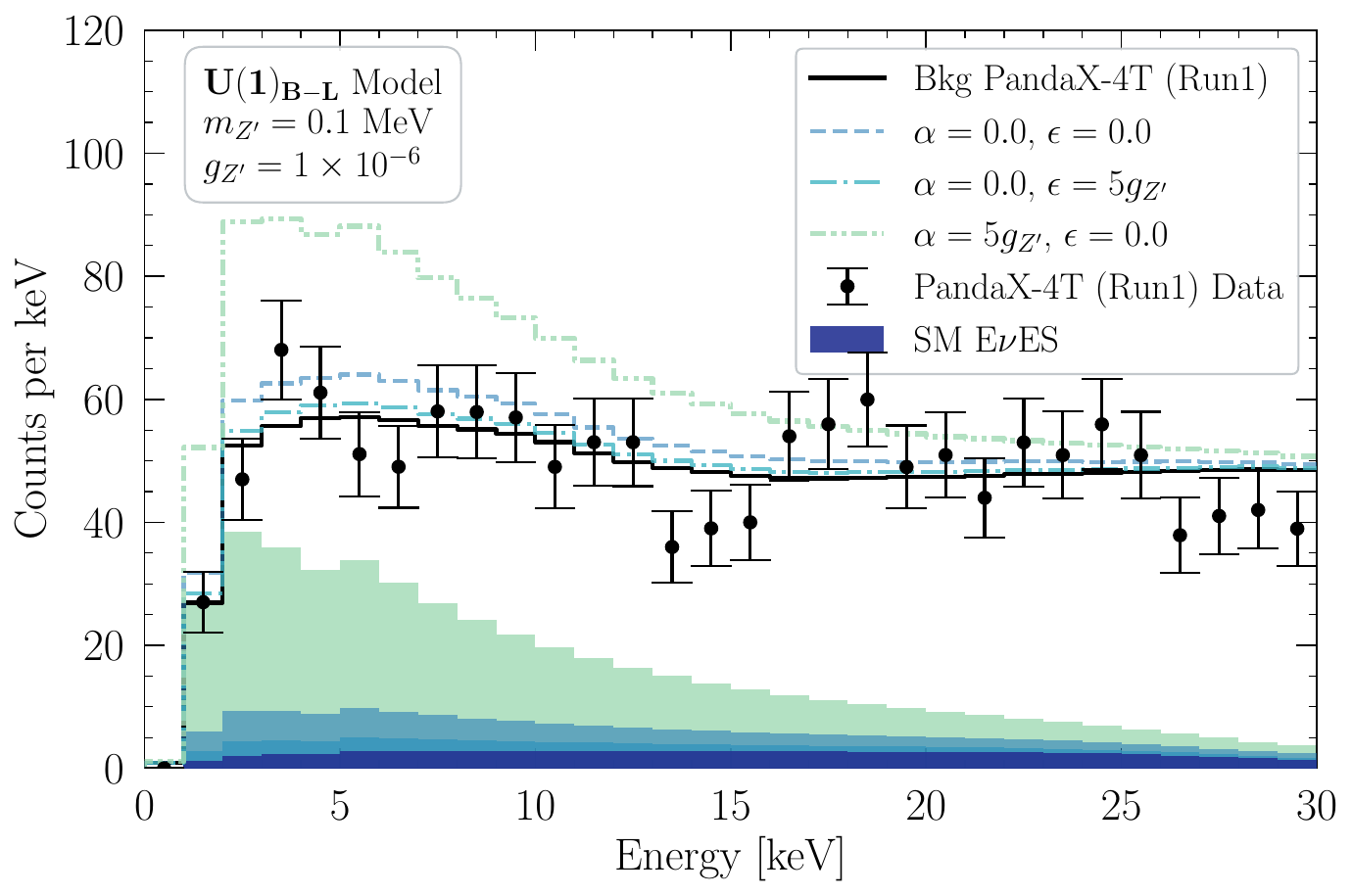}
	\\
	\hspace{7mm} (a) \hspace{82mm} (b)
	\includegraphics[scale=0.38]{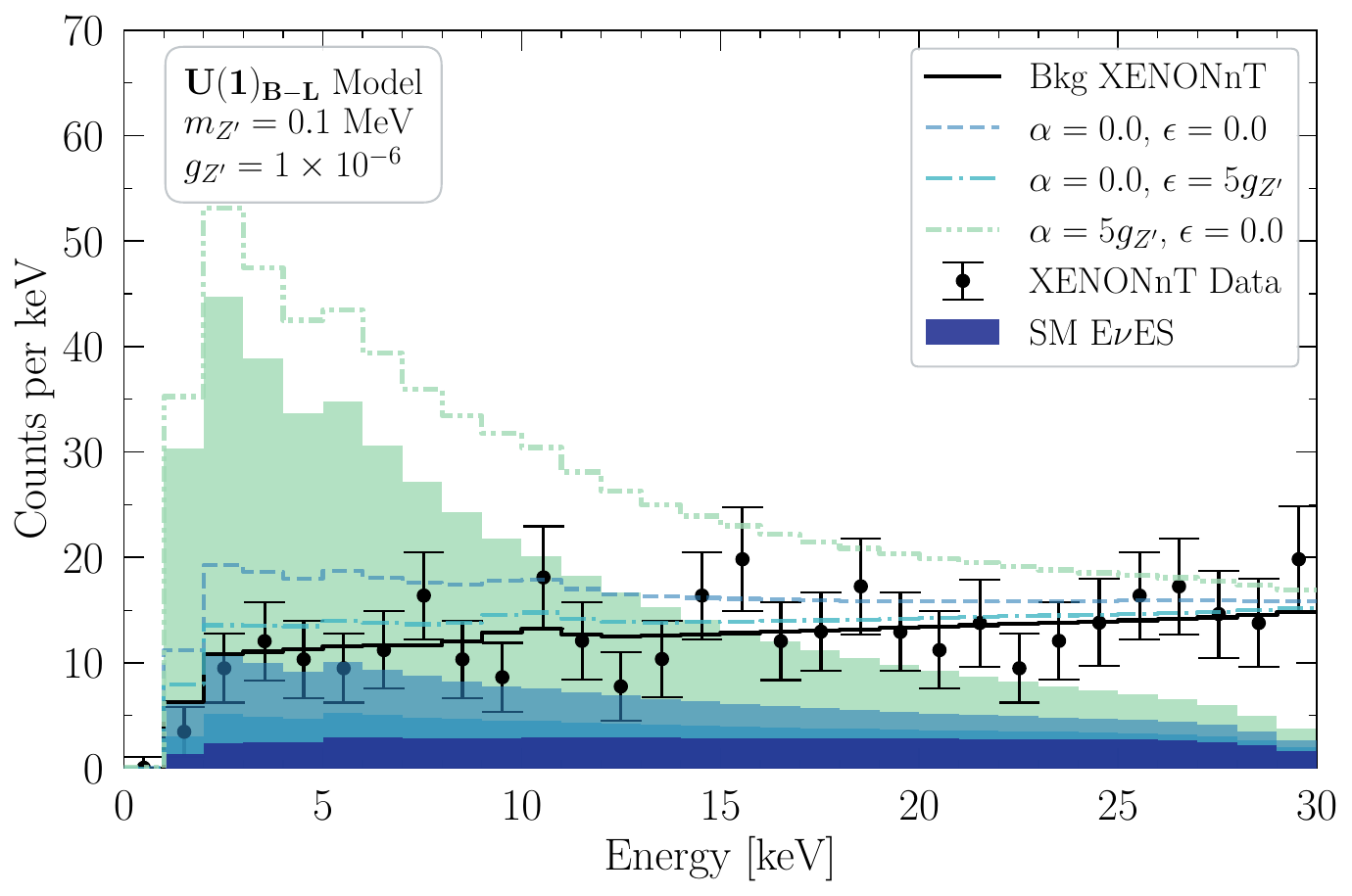}
	\\
	\hspace{7mm} (c)
	\caption{Predicted signals of $U(1)_{B-L}$ model as a function of electron recoil energy for the PandaX-4T (a) Run0 and (b) Run1,  and (c) XENONnT datasets.  Filled histograms denote individual E$\nu$ES signal contributions; outlined histograms of the same color show their respective contributions to the total background. The predicted events are evaluated for a benchmark mass $m_{Z'}=0.1$ MeV and various couplings as shown in the legend.}
	\label{fig:count_rate}
\end{figure*}
%

\begin{figure*}[ht!]
	\centering
	\includegraphics[scale=0.45]{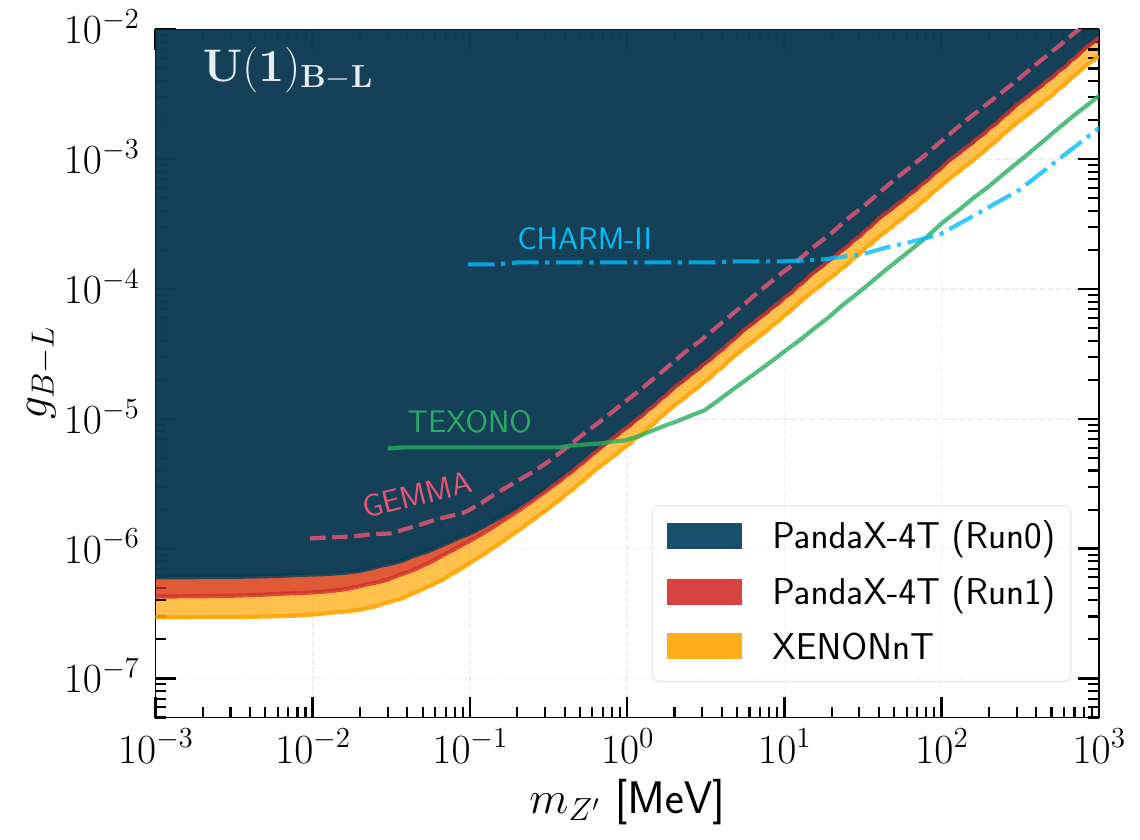}
	\includegraphics[scale=0.45]{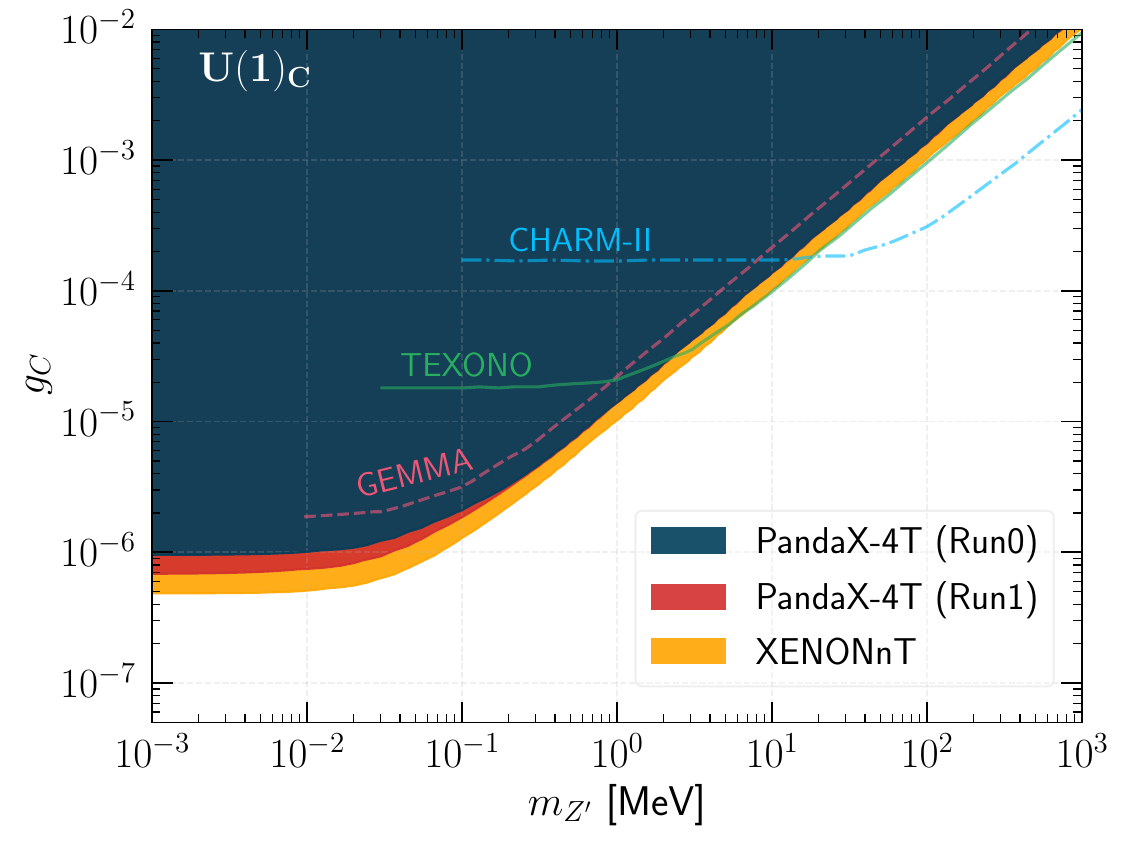} 
    \\
    (a) \hspace{8cm} (b)
    \\
	\includegraphics[scale=0.45]{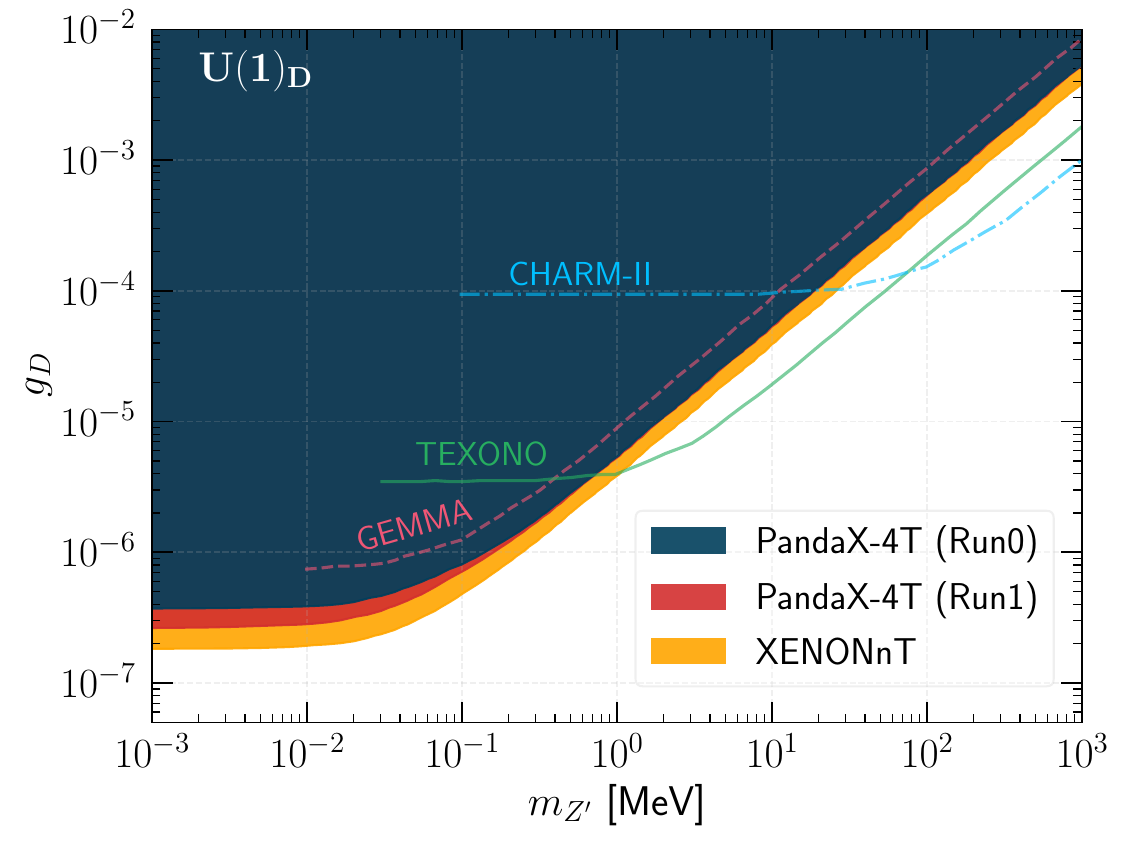}
	\includegraphics[scale=0.45]{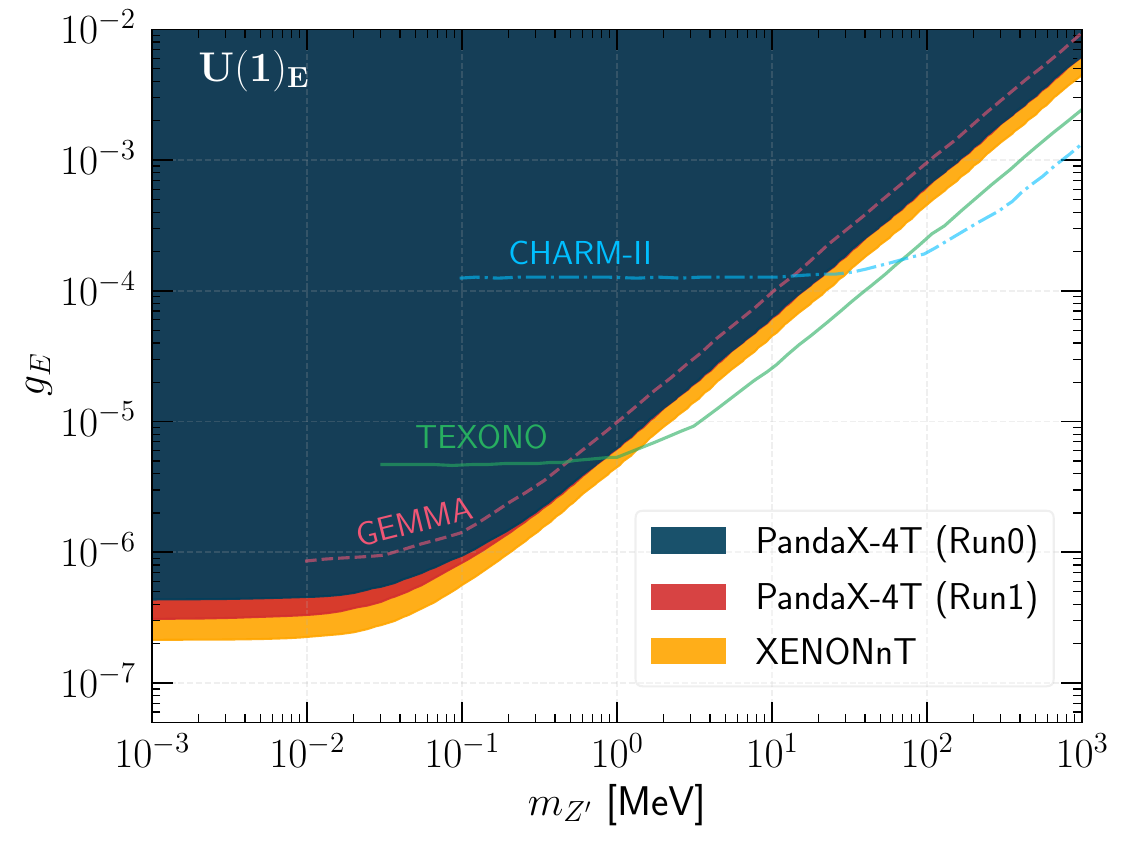}
    \\
    (c) \hspace{8cm} (d)
    \\
	\includegraphics[scale=0.45]{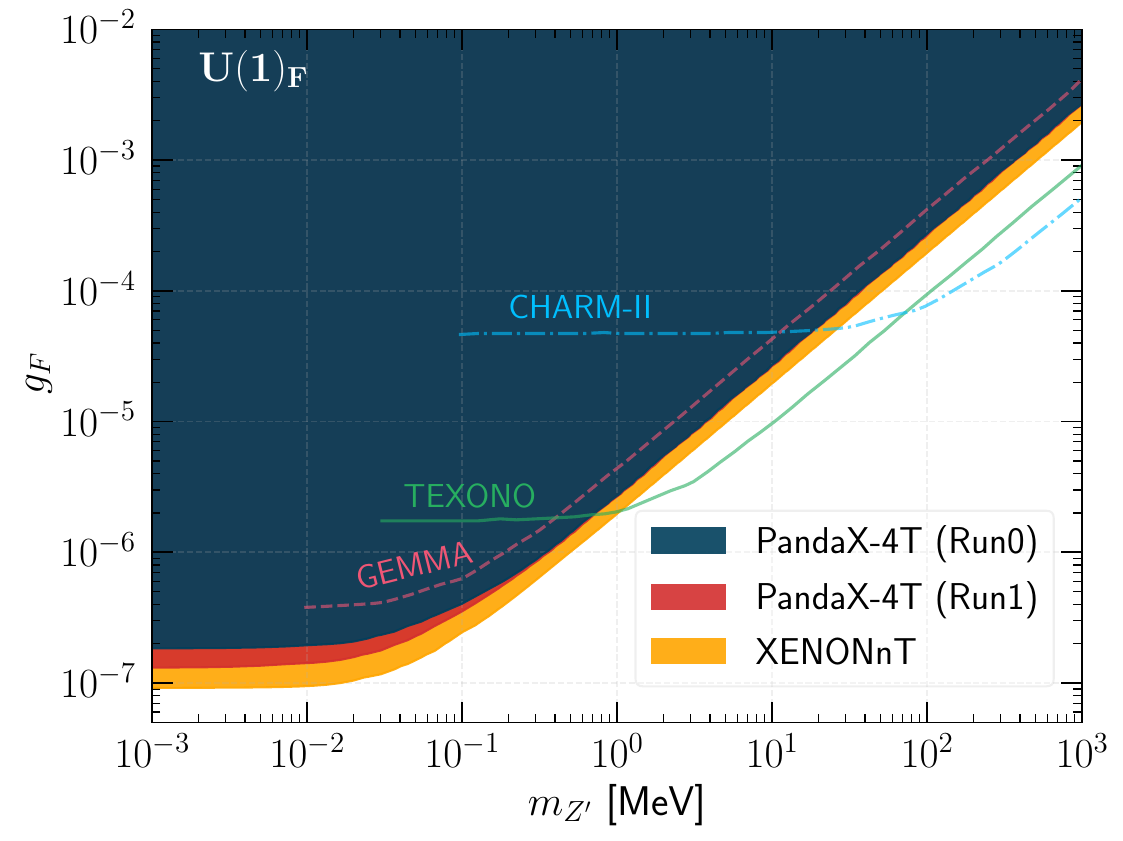}
	\includegraphics[scale=0.45]{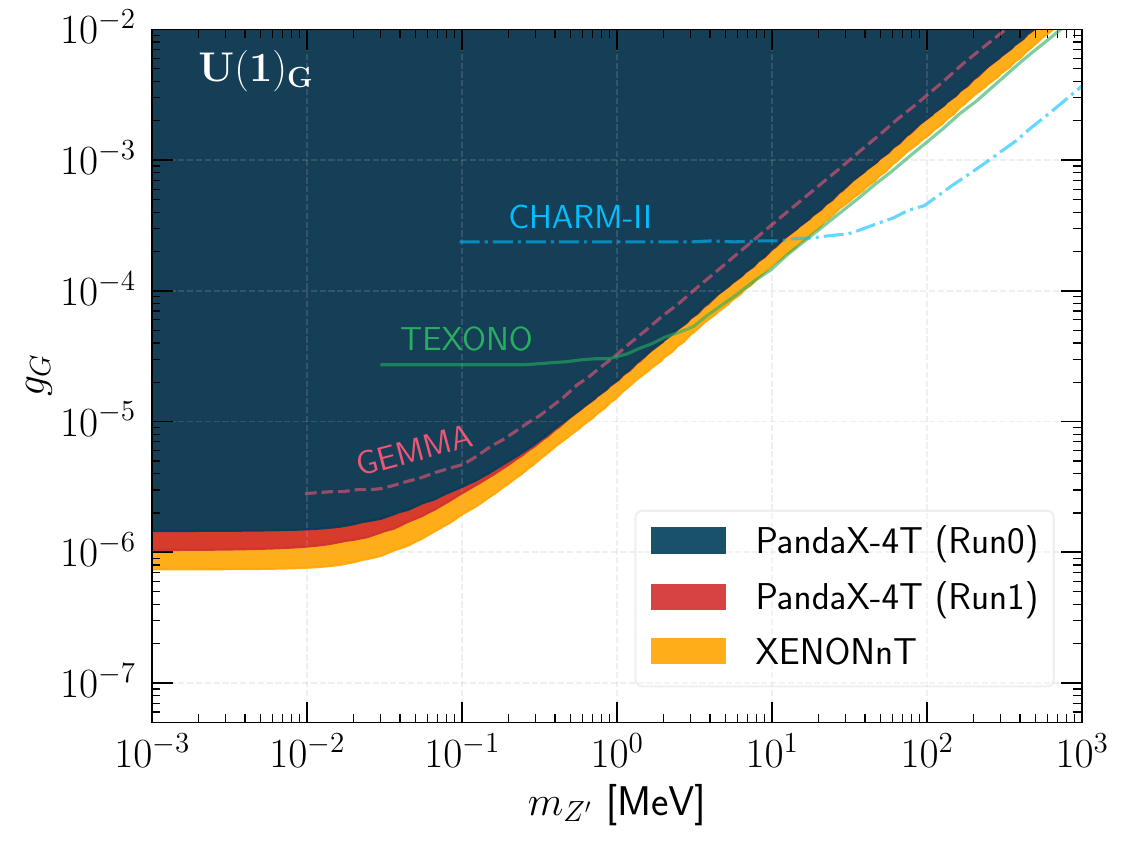}
    \\
    (e) \hspace{8cm} (f)
    \\
    \caption{The $90\%$ CL exclusion limits at 2 dof in the $(m_{Z'}, g_X)$ mass-coupling parameter space for the six considered $U(1)_X$ dark photon models. To isolate the pure gauge interactions, the mixing parameters are fixed to $\alpha=\epsilon=0$. The newly derived bounds from the recent PandaX-4T (Run0 and Run1) and XENONnT electronic recoil datasets are superimposed with existing experimental constraints. Previous limits obtained from the CHARM-II, TEXONO, and GEMMA experiments are adopted from Ref.~\cite{Lindner:2018kjo}.}
	\label{fig:limit_dp_models}
\end{figure*}

In Fig.~\ref{fig:count_rate}, we illustrate the predicted differential event rates as a function of the recoil energy of the PandaX-4T Run0 (a), PandaX-4T Run1 (b), and XENONnT (c) datasets. 
We choose the specific B-L scenario, since it is among the most popular $Z'$ scenarios in explaining the existence of new physics. The differential rates are consistently normalized in $\text{keV}^{-1}$.
The experimental data points, along with their corresponding uncertainties, are denoted by the black markers, while the solid black lines represent the total background models provided by the respective collaborations. The SM E$\nu$ES expected rates are depicted by the dark blue shaded regions, which consistently constitute a subdominant fraction of the total background across the entire energy range for all three datasets.
To illustrate the phenomenological impact of the $U(1)_{B-L}$ extension, we superimpose the expected BSM signal spectra for a benchmark $Z'$ mediator mass of $m_{Z'} = 0.1$~MeV and a gauge coupling of $g_{Z'} = 1 \times 10^{-6}$. We systematically investigate the effects of the mass mixing ($\alpha$) and kinetic mixing ($\epsilon$) parameters to highlight their distinct spectral contributions. The baseline BSM scenario with pure gauge interactions ($\alpha=0.0, \epsilon=0.0$) is denoted by the thin dashed cyan lines. Introducing a non-zero kinetic mixing ($\alpha=0.0, \epsilon=5g_{Z'}$, thick dashed cyan lines) leads to a moderate enhancement in the event rate at low energies. Conversely, turning on the mass mixing parameter ($\alpha=5g_{Z'}, \epsilon=0.0$, dash-dotted green lines) produces a drastic amplification of the differential rate, particularly at recoil energies below $10$~keV. As visually evident across all three panels, this pronounced low-energy excess significantly overshoots the observed experimental data points within the ROI.

\begin{table*}[ht]
	\caption{90\% CL (2 dof) upper limits on the coupling constants for each model under consideration. The table compares limits from PandaX-4T (Run0 and Run1) and XENONnT with existing bounds obtained from previous works, where limits from CHARM-II, TEXONO, and GEMMA are taken from Ref. \citep{Lindner:2018kjo}.  The reported values represent the most stringent upper limits within the considered parameter region. }
	\begin{center}
    	\label{tab:couplings_dp}
		\begin{ruledtabular}
		\begin{tabular}{ l c c c c c c}
            Experiments & $g_\text{B-L}$ & $g_C$ & $g_D$ & $g_E$ & $g_F$ & $g_G$ \\ 
			\hline
			\textbf{PandaX-4T (Run0)} & $\lesssim 5.99\times 10^{-7}$ & $\lesssim 9.52\times 10^{-7}$ & $\lesssim 3.71\times 10^{-7}$ & $\lesssim 4.35\times10^{-7}$ & $\lesssim 1.83\times 10^{-7}$ & $\lesssim 1.45\times 10^{-6}$
			\\ 
			\textbf{PandaX-4T (Run1)} & $\lesssim 4.26\times 10^{-7}$ & $\lesssim 6.83\times 10^{-7}$ & $ \lesssim 2.63\times 10^{-7}$ & $\lesssim 3.09\times10^{-7}$ & $\lesssim 1.31\times 10^{-7}$ & $\lesssim 1.04\times 10^{-6}$
			\\ 
			\textbf{XENONnT} & $\lesssim 2.95\times 10^{-7}$ & $\lesssim 4.83\times 10^{-7}$ & $ \lesssim 1.82\times 10^{-7}$ & $\lesssim 2.13\times10^{-7}$ &  $\lesssim 9.18\times 10^{-8}$ & $\lesssim 7.41\times 10^{-7}$
			\\ 
			CHARM-II \citep{Lindner:2018kjo}& $\lesssim 1.55\times10^{-4}$ & $\lesssim 1.72\times 10^{-4}$ & $ \lesssim 9.42 \times 10^{-5}$ & $\lesssim 1.25\times10^{-4}$ &  $\lesssim 4.63\times 10^{-5}$ & $\lesssim 2.37\times 10^{-4}$
			\\
			TEXONO \citep{Lindner:2018kjo}& $\lesssim 5.92\times10^{-6}$ & $\lesssim 1.81\times 10^{-5}$ & $ \lesssim 3.48 \times 10^{-6}$ & $\lesssim 4.70\times10^{-6}$ &  $\lesssim 1.74\times 10^{-6}$ & $\lesssim 2.72\times 10^{-5}$
			\\
			GEMMA \citep{Lindner:2018kjo}& $\lesssim 1.20\times10^{-6}$ & $\lesssim 1.87\times 10^{-6}$ & $ \lesssim 7.43\times 10^{-7}$ & $\lesssim 8.56\times10^{-7}$ &  $\lesssim 3.79\times 10^{-7}$ & $\lesssim 2.81\times 10^{-6}$
			\\
		\end{tabular}
		\end{ruledtabular}
	\end{center}
\end{table*}

We present the 90\% CL exclusion regions with 2 dof on the ($m_{Z'}$,$g_X$) parameter space for the $U(1)_\text{B-L}$, $U(1)_C$, $U(1)_D$, $U(1)_E$, $U(1)_F$, $U(1)_G$ dark photon models in Fig. \ref{fig:limit_dp_models}. 
We also tabulate our results as well as limits from reactor facilities in Table~\ref{tab:couplings_dp}, which can be extracted directly from these figures.
To isolate the pure gauge interaction contributions, we set the mixing parameters to $\alpha=\epsilon=0$. Our derived bounds from the recent PandaX-4T (Run0 and Run1) and XENONnT datasets are superimposed with existing constraints from accelerator and reactor neutrino experiments—specifically CHARM-II, TEXONO, and GEMMA—as derived in Ref.~\cite{Lindner:2018kjo}.

As depicted across all panels, the exclusion contours exhibit two distinct kinematic regimes. In the low-mass limit ($m_{Z'} \lesssim 0.1 $ MeV), the mediator mass becomes negligible compared to the typical momentum transfer ($q^2$) of the solar E$\nu$ES process. Consequently, the cross-section loses its $m_{Z'}$ dependence, leading to an asymptotic plateau in the upper bounds of the gauge coupling.
As quantified in Table~\ref{tab:couplings_dp}, in this asymptotic regime for the benchmark $U(1)_{B-L}$ model, for example, 
Run1 dataset of the PandaX-4T improves upon the Run0 by a factor of 1.41, yielding $g_\text{B-L}\lesssim 4.26 \times 10^{-7}$. Benefiting from its superior exposure and ultra-low background rejection, XENONnT provides the most stringent constraint globally at $g_\text{B-L}\lesssim 2.95 \times 10^{-7}$.
Conversely, in the heavy-mediator regime ($m_{Z'} \gtrsim 1$ MeV), the constraints weaken dynamically. Here, the momentum transfer becomes subdominant to the mediator mass, and the differential cross-section undergoes heavy-propagator suppression, scaling approximately as $1/m_{Z'}^4$. This kinematic penalty forces the exclusion limits to scale as $g_X \propto m_{Z'}$, which is visually evident as a constant positive slope in the log-log parameter space. Despite this suppression, our derived limits from the direct detection experiments completely envelop the parameter space previously probed by the GEMMA reactor experiment across all considered models. Furthermore, for $m_{Z'} \lesssim 10 $ MeV, the XENONnT and PandaX-4T constraints improve upon the TEXONO and CHARM-II bounds by more than one and three orders of magnitude, respectively. It is only in the significantly higher-mass region ($m_{Z'} > 10^2 $ MeV) that the high-energy accelerator data from CHARM-II begins to intersect with the solar neutrino limits due to the differing $q^2$ regimes.

A comparative analysis of the six $U(1)_X$ extensions reveals a distinct sensitivity hierarchy fundamentally rooted in their respective leptonic charge assignments. Because the magnitude of the BSM contribution to the E$\nu$ES signal is directly proportional to the effective lepton charges, models embedded with larger charge matrices predict significantly amplified recoil spectra. To remain consistent with the null-hypothesis observations from the respective detectors, models generating these intrinsically higher event rates must be subjected to proportionally tighter upper limits on their coupling constant $g_X$. Based on the asymptotic constraints summarized in Table~\ref{tab:couplings_dp}, we observe an exact bound sequence: $g_F < g_D < g_E <g_{B-L} < g_C < g_G $. This empirically derived hierarchy flawlessly mirrors the gauge charge assignments 
in Table~\ref{tab:dp_charges}. 
Notably, the $U(1)_F$ model—characterized by the largest anomaly-free leptonic charge configuration—experiences the most severe experimental restriction, driven down to $g_\text{F}\lesssim 9.18 \times 10^{-8}$ by the XENONnT dataset, whereas the $U(1)_G$ model undergoes the weakest bound ($g_\text{G}\lesssim 7.41 \times 10^{-7}$).

\begin{figure}[htb]
	\centering
	\includegraphics[scale=0.45]{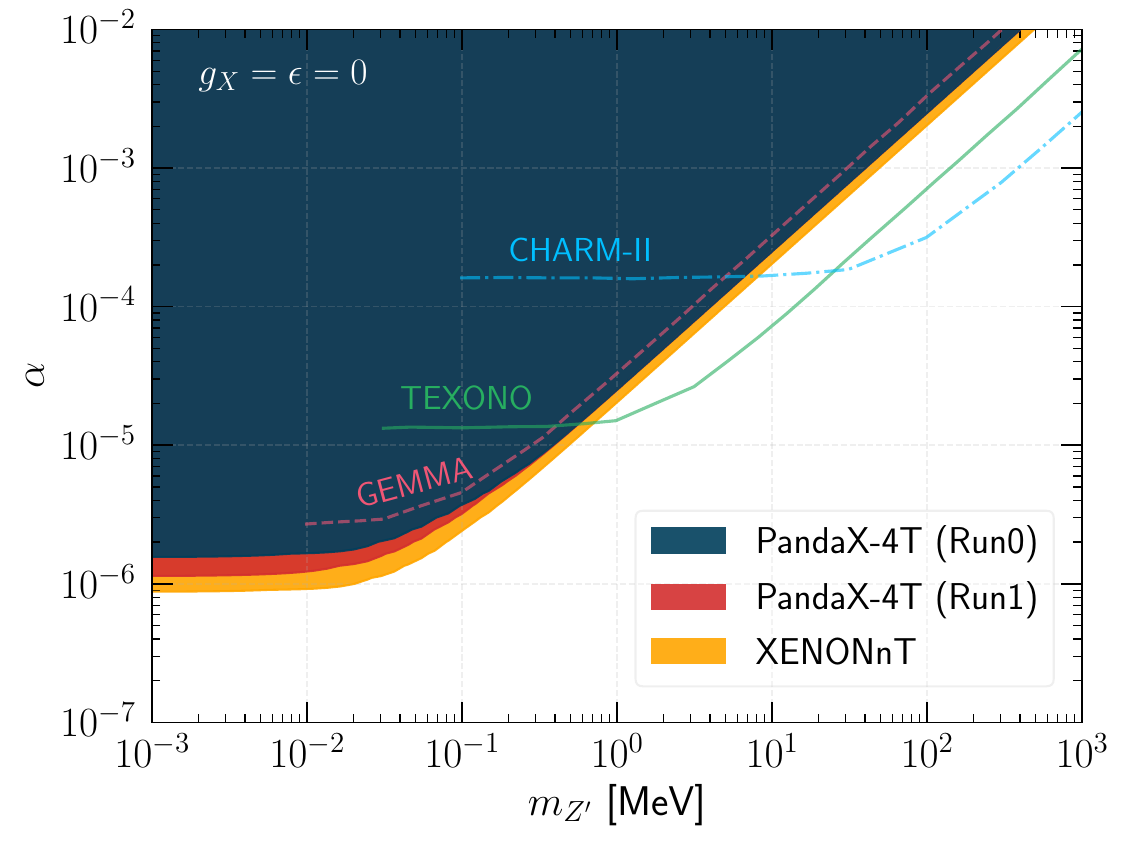}
    \caption{The $90\%$ CL exclusion limits at 2 dof in the $(m_{Z'}, \alpha)$ mass mixing parameter space, assuming the pure mass mixing scenario ($g_X=\epsilon=0$). Because the direct gauge coupling is turned off, these bounds are intrinsically independent of the specific $U(1)_X$ dark photon charge assignments. The newly derived constraints from the PandaX-4T (Run0 and Run1) and XENONnT datasets are superimposed with existing experimental limits. Previous bounds obtained from the CHARM-II, TEXONO, and GEMMA experiments are adopted from Ref.~\cite{Lindner:2018kjo}.}
	\label{fig:limit_alpha_mzp}
\end{figure}

\begin{figure*}[htb]
	\centering
	\includegraphics[scale=0.6]{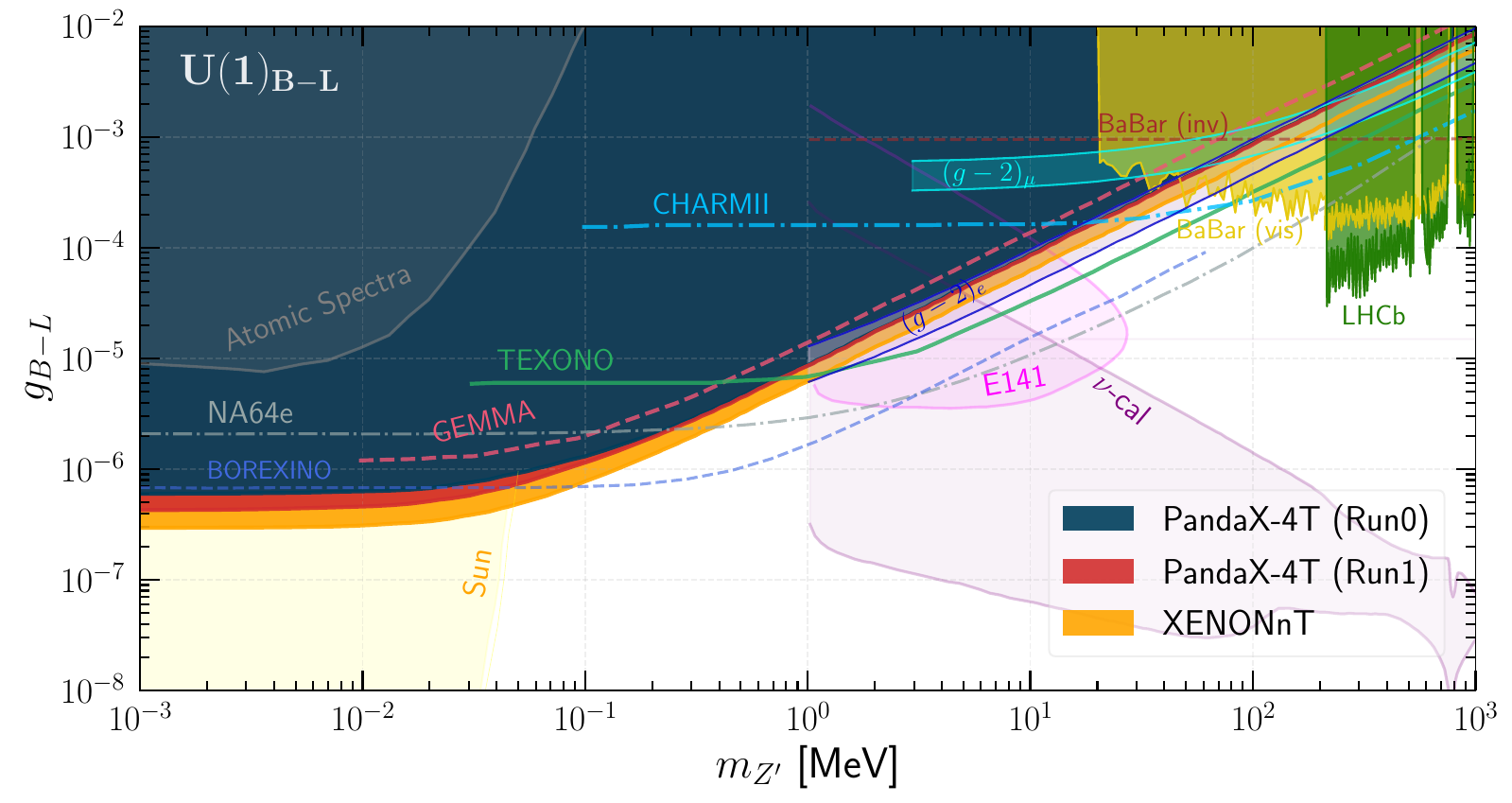}
	\\
    \caption{The $90\%$ CL exclusion limits at 2 dof in the $(m_{Z'}, g_{B-L})$ parameter space for the $U(1)_{B-L}$ dark photon model, assuming pure gauge interactions ($\alpha=\epsilon=0$). The newly derived bounds from the PandaX-4T (Run0 and Run1) and XENONnT datasets are superimposed with existing experimental and astrophysical constraints. Previous limits are adopted as follows: CHARM-II, TEXONO, and GEMMA from Ref.~\citep{Lindner:2018kjo}; BOREXINO from Ref.~\citep{Coloma:2022umy}; E141 and $\nu$-cal from Refs.~\citep{Asai:2022zxw,Asai:2023mzl}; NA64e from Ref.~\citep{NA64:2022yly}; $(g-2)_e (^{87}\mathrm{Rb})$ from Ref.~\citep{Asai:2022zxw}; $(g-2)_\mu$ from Ref.~\citep{AtzoriCorona:2022moj}; BaBar (visible and invisible) from Ref.~\citep{Asai:2023mzl}; LHCb from Ref.~\citep{Ilten:2018crw}; atomic spectra from Ref.~\citep{Bartlett:1988yy}; and the solar cooling bound (Sun) from Ref.~\citep{Li:2023vpv}.}
	\label{fig:comparison_gbl}
\end{figure*}

\begin{figure*}[htb]
	\centering
	\includegraphics[scale=0.63]{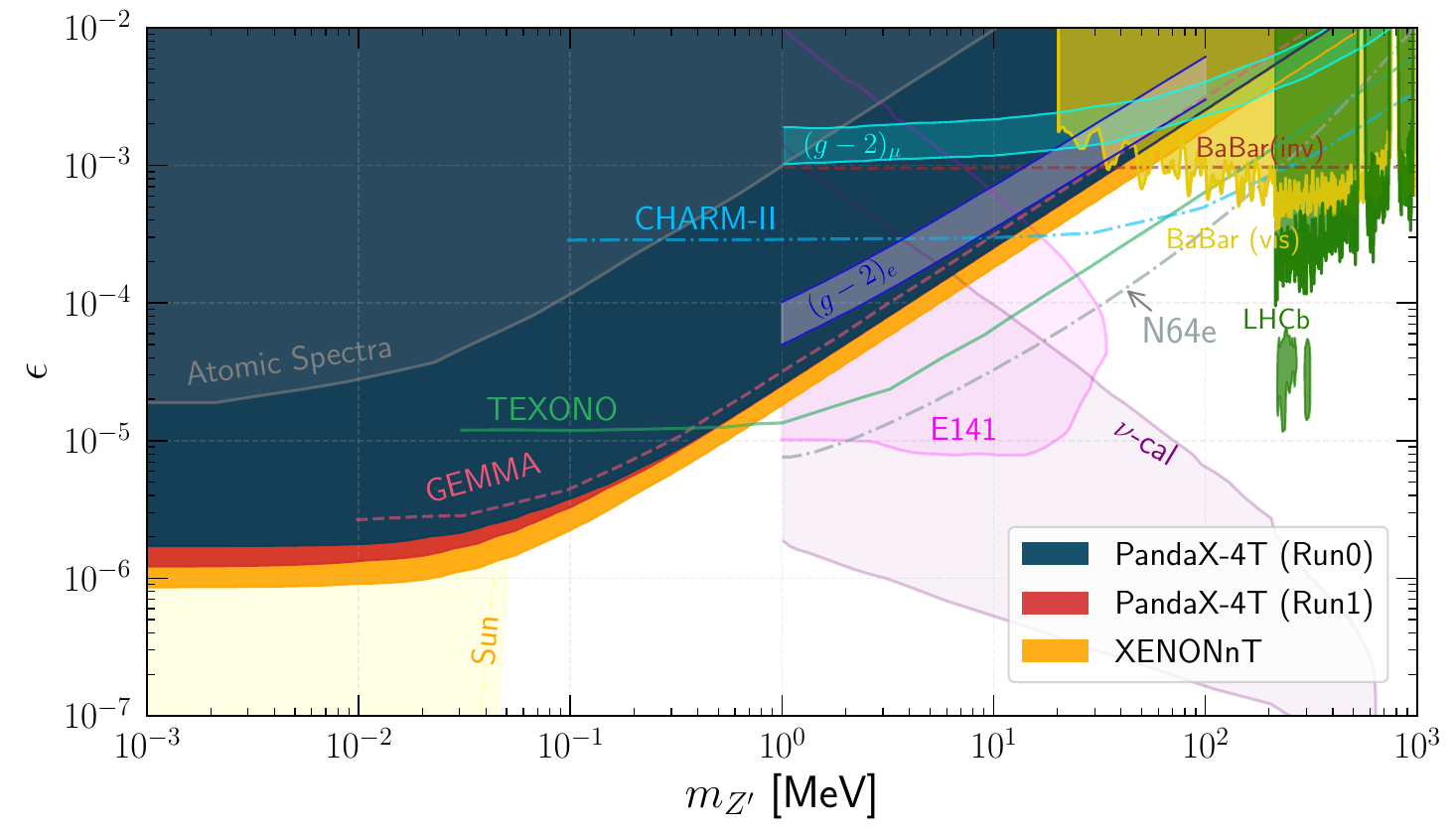}
	\\
    \caption{The $90\%$ CL exclusion limits at 2 dof in the $(m_{Z'}, \epsilon)$ mass-kinetic mixing parameter space, assuming the pure kinetic mixing scenario ($g_X=\alpha=0$). Because the direct gauge coupling is turned off, these bounds are intrinsically independent of the specific $U(1)_X$ dark photon charge assignments. The newly derived constraints from the PandaX-4T (Run0 and Run1) and XENONnT datasets are superimposed with a comprehensive compilation of previous limits. Existing bounds are adopted as follows: CHARM-II, TEXONO, and GEMMA from Ref.~\citep{Lindner:2018kjo}; E141 from Ref.~\citep{Bjorken:2009mm}; $\nu$-cal from Ref.~\citep{Blumlein:2013cua}; NA64 from Ref.~\citep{Banerjee:2019pds}; $(g-2)_e$ from Ref.~\citep{Morel:2020dww}; $(g-2)_\mu$ from Ref.~\citep{Tsai:2019buq}; BaBar (visible) from Ref.~\citep{BaBar:2014zli}; BaBar (invisible) from Ref.~\citep{BaBar:2017tiz}; LHCb from Ref.~\citep{Ilten:2018crw}; atomic spectra from Ref.~\citep{Jaeckel:2010xx}; and the solar cooling bound (Sun) from Ref.~\citep{Li:2023vpv}.}
	\label{fig:comparison_eps}
\end{figure*}
We next consider the orthogonal scenario in which $g_X=\epsilon=0$, enabling us to place exclusive constraints on the mass mixing parameter $\alpha$ and the $Z'$ mass. The 90\% CL exclusion regions with 2 dof derived from PandaX-4T (Run0 and Run1) and XENONnT datasets on the ($m_{Z'}$,$\alpha$) parameter space are shown in Fig~\ref{fig:limit_alpha_mzp}, superimposed with previous limits from CHARM-II, TEXONO, and GEMMA. Crucially, because the direct gauge coupling is completely turned off ($g_X=0$), the BSM interaction is governed solely by the mass mixing mechanism. Consequently, the limits obtained in this scenario are entirely independent of the specific $U(1)_X$ dark photon charge assignments, rendering them universally applicable (model-independent) across all theoretical frameworks considered in this study. Similar to the mass-coupling plane, the exclusion contours exhibit an asymptotic plateau in the low-mass regime ($m_{Z'} \lesssim 0.1 $ MeV) where the momentum transfer dominates over the mediator mass. In this kinematic region, we obtain an upper bound of $\alpha\lesssim 1.56\times10^{-6}$ from PandaX-4T (Run0) dataset. The increased exposure of PandaX-4T (Run1) tightens this bound to $\alpha\lesssim 1.14\times10^{-6}$. Consistently, the XENONnT dataset imposes the most rigorous constraint across the parameter space, yielding $\alpha\lesssim 8.75\times10^{-7}$. As evident in Fig.~\ref{fig:limit_alpha_mzp}, these newly derived limits from recent datasets constitute a substantial advancement over previous constraints. In the low-mass region, our results significantly improve upon the CHARM-II and TEXONO limits by multiple orders of magnitude while fully covering and superseding the entire parameter space previously excluded by the GEMMA reactor experiment. Ultimately, this specific evaluation is highly relevant for phenomenological models characterized by a highly suppressed dark photon coupling and vanishing kinetic mixing, where any new physics in the solar neutrino sector would be entirely driven by sizable mass mixing.

To contextualize the phenomenological impact of our findings, we place our newly derived limits for the $U(1)_\text{B-L}$ framework into a broader global landscape of dark photon searches. In Fig.~\ref{fig:comparison_gbl}, the 90\% CL exclusion limits obtained from the PandaX-4T and XENONnT datasets on the ($m_{Z'}$,$g_\text{B-L}$) parameter space are superimposed with existing constraints from a diverse array of experimental and astrophysical probes. In the 
low mediator mass regime ($m_{Z'} \lesssim 0.1 $ MeV), our bounds establish strict dominance over terrestrial limits. The direct detection constraints completely cover the exclusion zones previously set by Atomic Spectra and the NA64e experiment, pushing deep enough into the parameter space to seamlessly connect with the stringent Sun limit \footnote{It is worth mentioning that stellar cooling bounds derived from horizontal branch stars and red giants ~\citep{Li:2023vpv}, though subject to larger astrophysical uncertainties, are generally more stringent than those derived from the Sun.}.
In this sub-MeV region, the ultra-low energy thresholds of LXe TPCs allow our constraints to surpass the highly competitive bounds derived from the Borexino solar neutrino measurements. As the mediator mass increases through the intermediate regime ($0.1\lesssim m_{Z'}\lesssim 100$ MeV), our bounds systematically surpass several established phenomenological territories. While the Borexino limit appears to be more stringent than the direct detection bounds in the higher mass range, our results supersede legacy limits derived from other neutrino-electron scattering experiments and rigorously probe previously unconstrained regions, partially covering the exclusion bounds established by beam dump experiments such as E141 and $\nu$-cal.
Furthermore, our derived limits intersect the parameter bands favored by $(g-2)_e$ and $(g-2)_\mu$. 
While our constraints exclude a substantial portion of these preferred regions, they do not entirely rule them out, leaving a residual allowed parameter space.
Finally, in the higher mass region extending toward the 1 GeV scale,
our limits exhibit strong complementarity with direct searches conducted at high-energy colliders. Although heavy-propagator suppression naturally weakens the direct detection limits in this kinematic regime, our results remain phenomenologically relevant, effectively bridging the gap and smoothly approaching the robust exclusion boundaries set by the BaBar (visible and invisible decay channels) and LHCb collaborations.

\begin{figure*}[htb]
	\centering
	\includegraphics[scale=0.29]{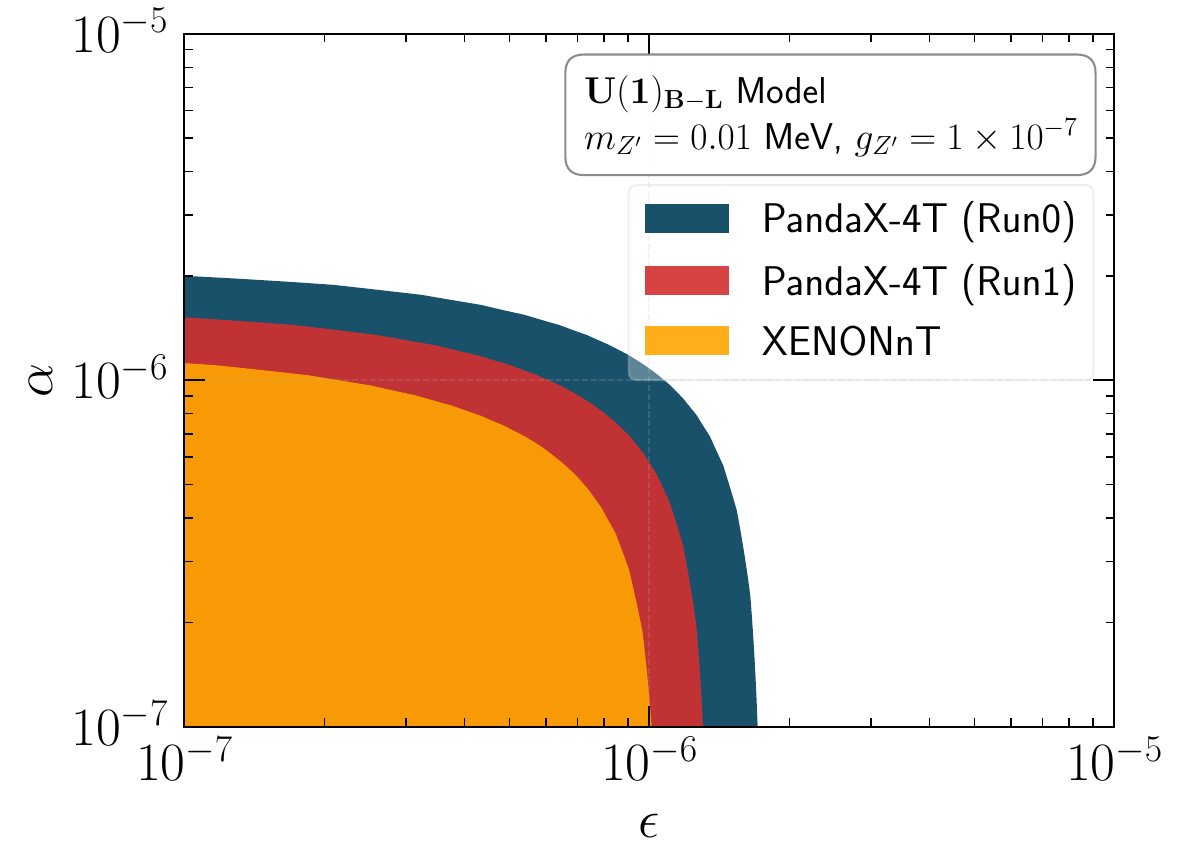}
	\includegraphics[scale=0.29]{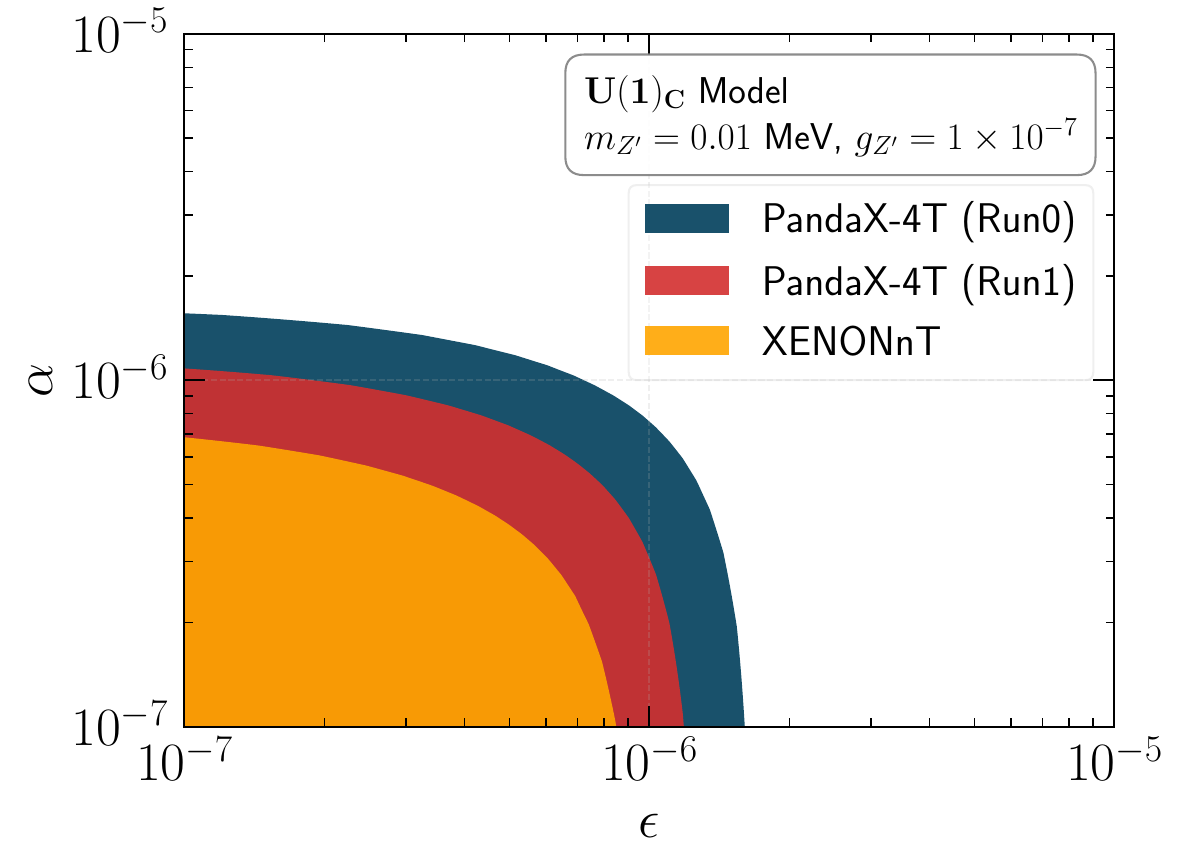}
		\includegraphics[scale=0.29]{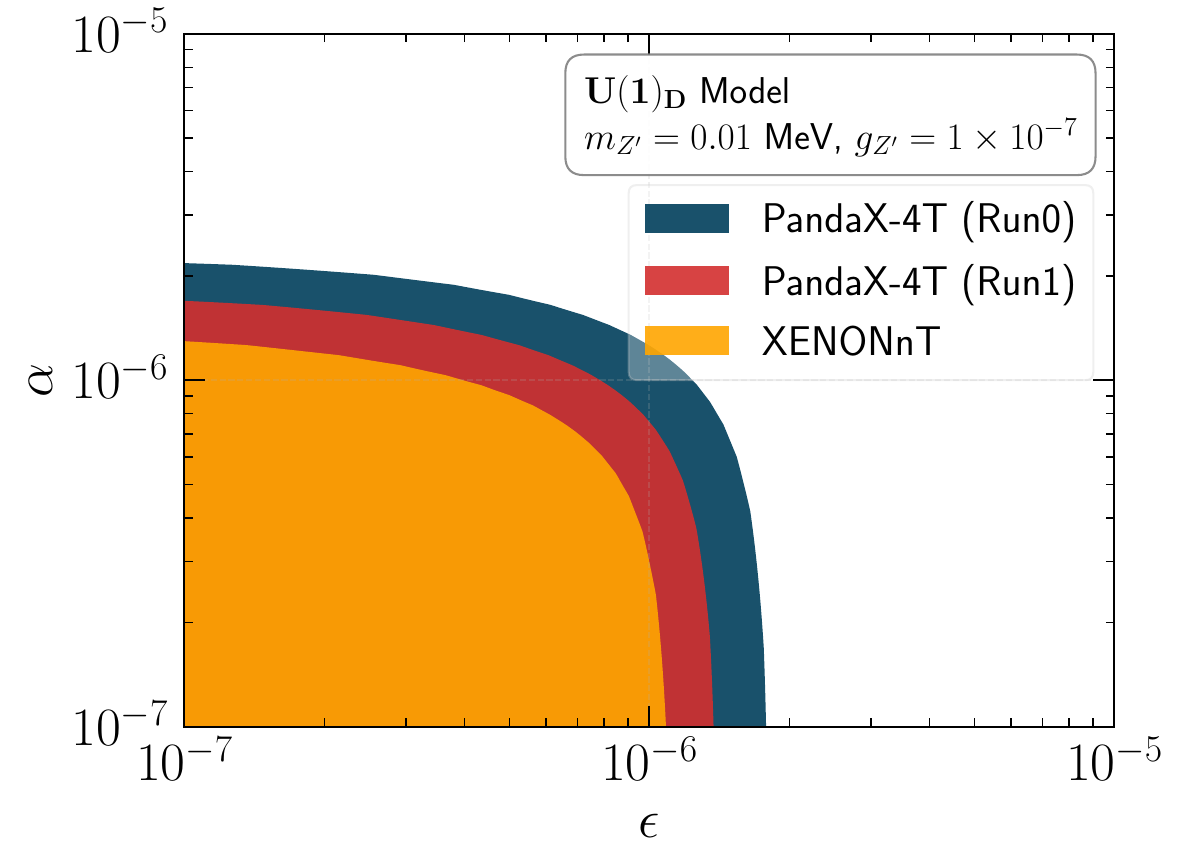}
		\includegraphics[scale=0.29]{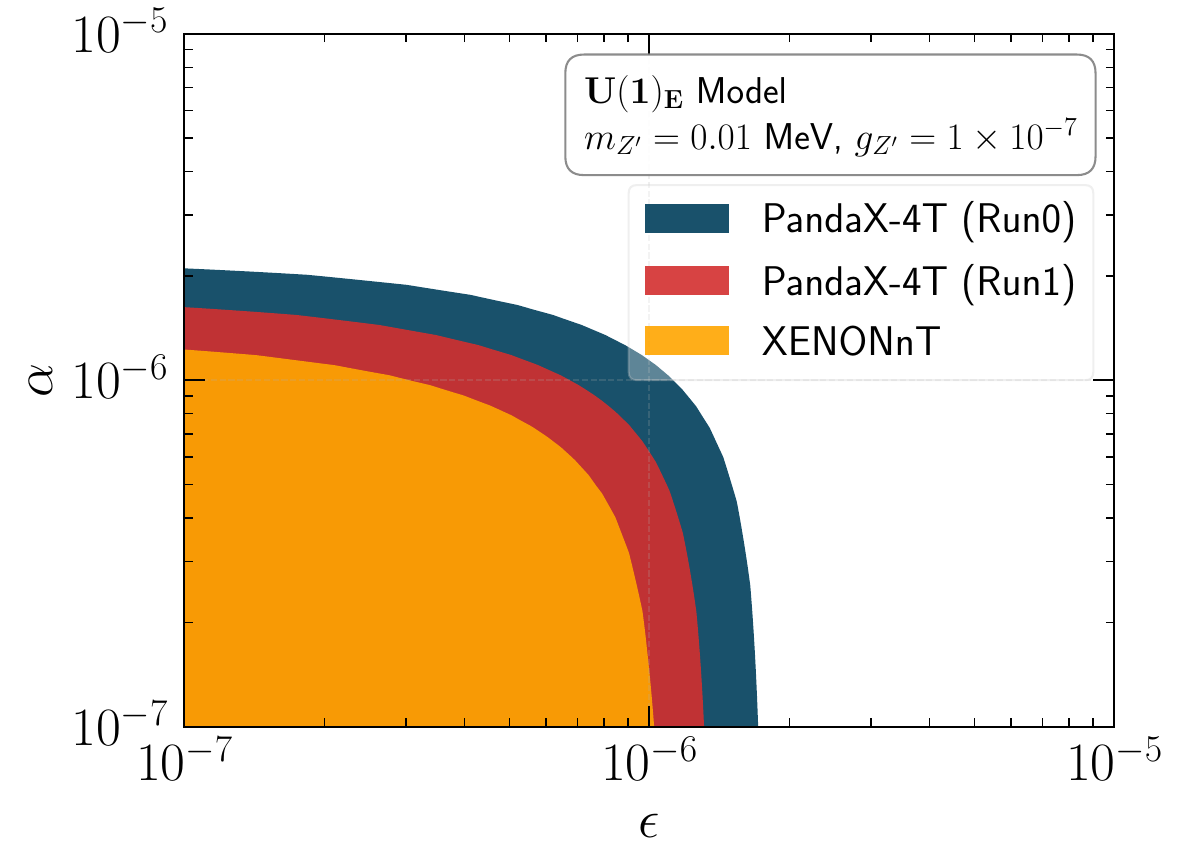}
		\includegraphics[scale=0.29]{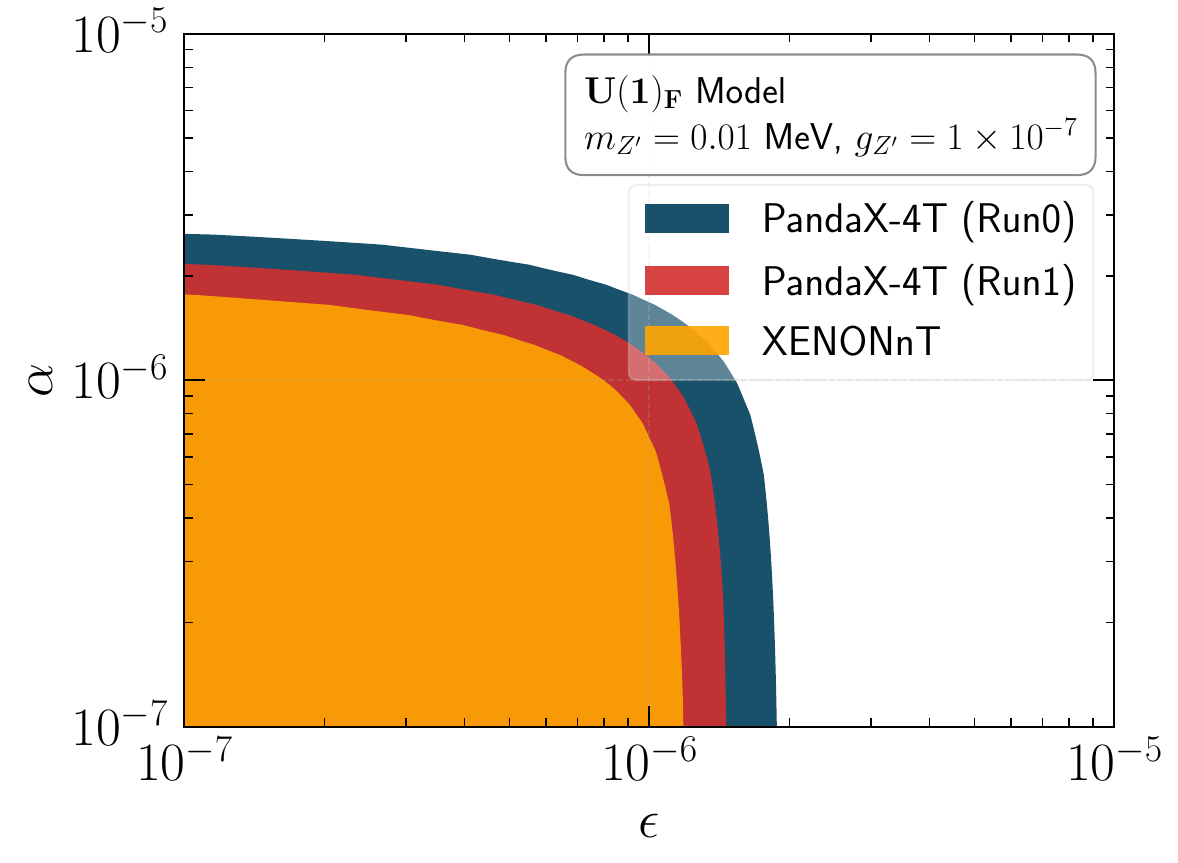}
		\includegraphics[scale=0.29]{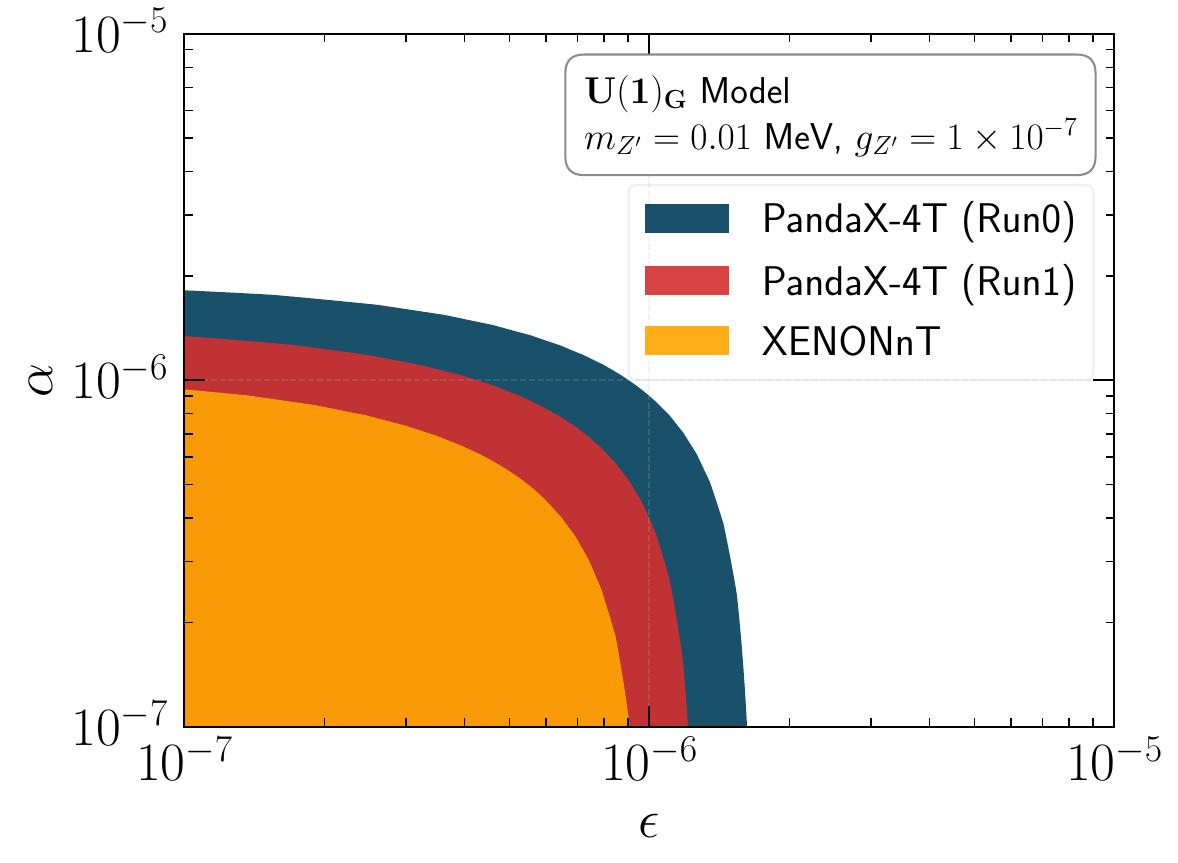}
    \caption{The $90\%$ CL allowed regions at 2 dof in the $(\epsilon, \alpha)$ parameter space for the six considered $U(1)_X$ dark photon models. The bounds are derived from the PandaX-4T (Run0 and Run1) and XENONnT electronic recoil datasets, evaluated at a fixed mediator mass of $m_{Z'} = 0.01\text{ MeV}$ and a baseline gauge coupling of $g_{Z'} = 1 \times 10^{-7}$. The shaded contours illustrate the simultaneously permitted values for the kinetic mixing ($\epsilon$) and mass mixing ($\alpha$) parameters across the different anomaly-free charge assignments.}
	\label{fig:limit_alpha_epsilon}
\end{figure*}

Subsequently, we investigate the pure kinetic mixing scenario by setting  $g_X=\alpha=0$, which enables us to place robust 90\% CL exclusion limits on the ($m_{Z'},\epsilon$) parameter space. The derived bounds from the PandaX-4T (Run0 and Run1) and XENONnT datasets are presented in Fig.~\ref{fig:comparison_eps}, superimposed with a comprehensive compilation of constraints from previous experimental efforts. 
Because the direct dark gauge coupling is strictly assumed to vanish in this scenario, the interactions are mediated exclusively through the kinetic mixing between the SM photon and the $Z'$ boson. Consequently, the derived exclusion limits are intrinsically independent of the specific $U(1)_X$ charge assignments, rendering them broadly applicable across various dark photon models.
In the low-mass asymptotic regime ($m_{Z'} \lesssim 0.1 $ MeV), we extract upper limits of $\epsilon\lesssim1.70 \times10^{-6}$ from PandaX-4T (Run0) and $\epsilon\lesssim1.21 \times10^{-6}$ from PandaX-4T (Run1). XENONnT establishes the most rigorous upper bound among the direct detection datasets, yielding  $\epsilon\lesssim8.56 \times10^{-7}$. As illustrated in Fig.~\ref{fig:comparison_eps}, the exclusion contours for $\epsilon$ exhibit a kinematic topology analogous to the mass-gauge coupling plane of the $U(1)_{B-L}$ model, characterized by a flat plateau at low mediator masses and a quadratic relaxation at higher masses due to heavy-propagator suppression. We tabulate these numerical limits, alongside the existing bounds from reactor facilities (CHARM-II, TEXONO, and GEMMA), in Table~\ref{tab:couplings_dp_eps}. A critical feature of our results is their significant intersection with the parameter regions favored by the anomalous magnetic dipole moments of the lepton sector. Our newly derived direct detection limits completely rule out the entire parameter space favored by the electron anomaly, $(g-2)_e$. Furthermore, they robustly constrain and exclude a substantial portion of the parameter band favored by the muon anomaly, $(g-2)_\mu$, leaving only a marginal unconstrained window at higher masses.

When contextualized within the broader experimental landscape, the solar neutrino limits provide vital complementarity. In the sub-MeV mass region, while the astrophysical stellar cooling limits remain the most stringent phenomenological bounds, our direct detection constraints decisively surpass the experimental limits from atomic spectra as well as legacy reactor experiments like TEXONO and GEMMA. At higher masses ($m_{Z'} \gtrsim 1 $ MeV), dedicated beam-dump and collider experiments—such as E141, NA64e, BaBar, and LHCb—maintain leading sensitivities. Nevertheless, the independent cross-check provided by PandaX-4T and XENONnT via entirely distinct physical processes (solar neutrino-electron elastic scattering) is phenomenologically invaluable. Ultimately, these results highlight that ton-scale direct detection experiments, characterized by their ultra-low recoil energy thresholds, massive target volumes, and continuously improving exposures, are highly competitive multi-purpose facilities capable of pushing the frontiers of dark sector physics.

\begin{table}[htpb]
	\centering
	\caption{90\% CL (2 dof) upper limits on the kinetic mixing parameter. The table compares our limits from PandaX-4T (Run0 and Run1) and XENONnT with existing limits from CHARM-II, TEXONO, and GEMMA derived in Ref. \citep{Lindner:2018kjo}.}
    \label{tab:couplings_dp_eps}
	\begin{ruledtabular}
    \begin{tabular}{lc}
		Experiments & \boldmath{$\epsilon$} \\
		\midrule
		\textbf{PandaX-4T (Run0)} & $\lesssim 1.70\times 10^{-6}$  \\
		\textbf{PandaX-4T (Run1)} & $\lesssim 1.21\times 10^{-6}$ \\
        \textbf{XENONnT} & $\lesssim 8.56\times 10^{-7}$ \\
        CHARM-II \citep{Lindner:2018kjo} & $\lesssim 1.61\times 10^{-4}$ \\
        TEXONO \citep{Lindner:2018kjo} & $\lesssim 1.33\times 10^{-5}$ \\
        GEMMA \citep{Lindner:2018kjo} & $\lesssim 2.70\times 10^{-6}$ \\
	\end{tabular}
    \end{ruledtabular}
\end{table}

As a final comprehensive scenario, we explore the simultaneous constraints on the mass mixing and kinetic mixing parameters. Figure~\ref{fig:limit_alpha_epsilon} illustrates the 90\% CL allowed regions with 2 dof in the ($\epsilon,\alpha$) parameter spaces for each $U(1)_X$ dark photon model. For this specific evaluation, we establish a fixed benchmark by setting the mediator mass to  $m_{Z'}=0.01$ MeV and the gauge coupling constant to $g_{Z'}=1\times10^{-7}$. Because both mixing mechanisms independently contribute to the total BSM event rate within the electron recoil spectra, their permitted values are strongly anti-correlated along the exclusion boundaries. As depicted by the contour shapes, accommodating a sizable kinetic mixing necessitates a highly suppressed mass mixing, and vice versa, to ensure the total differential cross-section remains within the experimentally permitted limits.
Furthermore, setting the gauge coupling to a non-zero constant introduces a model-dependent baseline interaction that is governed by the respective $U(1)_X$ leptonic charges. This pre-existing baseline inherently dictates the remaining "room" in the parameter space available for the mixing mechanisms. Consequently, we observe a distinct stringency hierarchy among the allowed contour areas that strongly depends on the specific dark photon charge assignments. In this configuration, the $U(1)_C$ model is subjected to the most stringent constraints (exhibiting the smallest allowed region), followed systematically by the $U(1)_G$, $U(1)_{B-L}$, $U(1)_E$, $U(1)_D$, and finally the $U(1)_F$ model, which possesses the most relaxed bounds. Consistent with the behavioral trends established in our previous single-parameter evaluations, the direct detection datasets display a clear hierarchy of experimental sensitivity. The XENONnT dataset consistently defines the innermost bounds, providing the most rigorous restrictions on the combined ($\epsilon,\alpha$) parameter space. This is followed by the PandaX-4T (Run1) data, which offers a visible enhancement over the PandaX-4T (Run0) constraints, reaffirming the profound impact of increased target exposure and optimized background discrimination in ton-scale liquid xenon experiments.

\section{Summary and Conclusions}\label{sec:summ}
In this work, we have systematically investigated the phenomenological prospects of probing anomaly-free dark photon models through solar neutrino-electron elastic scattering data. Utilizing recent low-energy electronic recoil datasets from ton-scale direct dark matter detection experiments PandaX-4T (Run0 and Run1) and XENONnT, we have derived robust constraints on the dark photon model parameter space. Our theoretical framework incorporated six distinct $U(1)_X$ gauge extensions—namely $U(1)_\text{B-L}, U(1)_C, U(1)_D, U(1)_E, U(1)_F, U(1)_G$. 

To comprehensively capture the new physics signatures, we have presented generalized cross-section formulations in terms of chiral couplings that encompass the mediator mass, gauge coupling, kinetic mixing, and mass mixing parameters.
By evaluating the solar neutrino fluxes with recent neutrino oscillation parameters, we have established rigorous 90\% CL exclusion limits across multiple parameter planes: ($m_{Z'},g_X$), ($m_{Z'}, \alpha$), ($m_{Z'},\epsilon$) and the correlated ($\epsilon,\alpha$) space. For the pure gauge interaction scenario, we have demonstrated that the constraint stringency inherently follows the magnitude of the theoretical leptonic charge assignments, with the $U(1)_F$ model yielding the most severe bounds. Furthermore, by isolating the pure mass mixing and pure kinetic mixing scenarios, we have derived model-independent constraints that are universally applicable across all considered $U(1)_X$ charge formulations.

Our derived limits significantly advance the current parameter landscape of the dark photon models. 
In the sub-MeV mediator mass regime, our results extracted from XENONnT and PandaX-4T datasets supersede existing constraints derived from accelerator and reactor neutrino facilities—such as CHARM-II, TEXONO, and GEMMA—by several orders of magnitude. Additionally, our limits decisively probe and exclude substantial portions of the parameter space favored by the electron and muon anomalous magnetic dipole moments. While astrophysical stellar cooling bounds and high-energy collider data maintain leading sensitivities in certain kinematic windows, our direct detection results offer an invaluable, independent probe via an entirely distinct physical channel. 
Future statistical improvements and further experimental ventures will be critical for refining these observations and probing dark photon model parameters with even greater accuracy.

Ultimately, our work underscores the profound capability of ultra-low energy threshold, ton-scale xenon-based dark matter detectors to serve as multi-purpose observatories. By leveraging electronic recoil signals induced by solar neutrinos, these facilities provide highly competitive and complementary probes of dark portals, particularly demonstrating exceptional sensitivity in the low-mass regions of dark photon models. 
We anticipate that the comprehensive limits and theoretical formulations presented herein will serve as a valuable resource for both the phenomenological and experimental communities in guiding future searches for physics beyond the SM.

\section*{Acknowledgments}
This work was supported by the Scientific and Technological Research Council of Türkiye (TUBITAK) under the project no: 124F416. M. Demirci thanks the CERN Theoretical Physics Department for its warm hospitality and support. We would like to acknowledge networking support of the COST Action CA21106 - COSMIC WISPers in the Dark Universe: Theory, astrophysics and experiments (CosmicWISPers). 

\appendix

\section{Anomaly-free Cancellation}\label{app:anomaly_free}
A chiral field $\psi^L_f$ or $\psi^R_f$ undergoes local transformation as
\begin{align}
\begin{split}
&\psi^L_f \to {\psi'}^L_f = e^{iQ^L_{Xf}\alpha(x)} \psi^L_f\\
&\psi^R_f \to {\psi'}^R_f = e^{iQ^R_{Xf}\alpha(x)} \psi^R_f.
\end{split}
\end{align}
where $f=\ell, q, e, u, d, h_1, h_2$ respectively denotes the index for left-handed lepton and quark, right-handed electron, up quark, and down quark, as well as the Higgs content $\Phi_1$ and $\Phi_2$. The field charge under $U(1)_X$ is denoted by $Q^{R,L}_{f}$. The scenario we consider focuses on the Type-I 2HDM. Fermions couple with $\Phi_2$ only in this model. 
To remain invariant, the following conditions need to be satisfied
\begin{align}
\begin{split}
Q^R_{Xd} - Q^L_{Xq} + Q_{Xh_2}=&0\\
Q^R_{Xu} - Q^L_{Xq} - Q_{Xh_2}=&0\\
Q^R_{Xe} - Q^L_{X\ell} + Q_{Xh_2}=&0.
\end{split}
\end{align}
Apart from these constraints, the model has to obey anomaly-free conditions. 
From $[SU(2)_L]^2 U(1)_X$ combination, denoting the charge of $U(1)_X$ as $X_L=Q^L_{Xf}$ or $X_R=Q^R_{Xf}$, we have $0 - \text{Tr}\left[\left\{\sigma^a/2,\sigma^b/2\right\} X_L\right]\propto - \sum X_L = - (2Q^L_{X\ell} + 3\times 2Q^L_{Xq}) = 0 $, so that 
\begin{align}
Q^L_{X\ell} = -3 Q^L_{Xq}.
\end{align}
With this result, the charges are constrained by
\begin{align}
\begin{split}
&Q^L_{Xq} = \frac{1}{2} (Q^R_{Xu} + Q^R_{Xd})\\
&Q^L_{X\ell} = -\frac{3}{2} (Q^R_{Xu} + Q^R_{Xd})\\
&Q^R_{Xe} = -(2Q^R_{Xu} + Q^R_{Xd}) \\
&Q_{h_2} = \frac{1}{2} (Q^R_{Xu} - Q^R_{Xd})
\end{split}
\end{align}
As for the case of $B-L$ model with $Q^R_{Xu}=Q^R_{Xd}=1/3$, then
\begin{align}
\begin{split}
&Q^L_{X\ell} = -\frac{3}{2}\left(\frac{1}{3} + \frac{1}{3}\right) = -1 \quad \text{and}\\
&Q^R_{Xe} = -\left(2\frac{1}{3} + \frac{1}{3}\right) = -1,
\end{split}
\end{align}
which is the second column in Table~\ref{tab:dp_charges}. We use this prescription for the other models, adapted from Ref.\citep{Campos:2017dgc}.

To complete the discussion, we note that right-handed neutrinos emerge naturally in this model. We can see from the $[U(1)_X]^3$ combination, we obtain $\text{Tr}\left[\left\{X_R, X_R\right\} X_R\right] - \text{Tr}\left[\left\{X_L, X_L\right\} X_L\right] \propto \sum X_R- \sum X_L \propto (Q_e^3 + 3Q_u^3 + 3Q_d^3) - (2Q_\ell^3 + 3\times 2Q_q^3) = 0 $, so that
\begin{align}
(Q^R_{Xe})^3 + 3(Q^R_{Xu})^3 + 3(Q^L_{Xd})^3 - 2(Q^L_{X\ell})^3 - 6(Q^L_{Xq})^3 = 0.
\end{align}
This anomaly condition is not cancelled by the current particle content. Adding right-handed neutrinos  with charge
\begin{align}
Q^R_{X\nu} = -(Q^R_{Xu} + 2Q^R_{Xd}),
\end{align}
is a straightforward solution to cancel this anomaly.

\end{document}